\documentclass[review,onefignum,onetabnum]{article}

\usepackage{latexsym}
\usepackage{amssymb,amsbsy,amsmath,amsfonts,amssymb,amscd,amsthm}
\usepackage{subfigure}
\usepackage{graphicx}
\usepackage{hyperref}
\usepackage{dsfont}
\usepackage{mathtools}
\usepackage{pdfsync}
\usepackage{epsfig,epstopdf}
\usepackage{caption}
\usepackage{xcolor}
\usepackage{hyperref}

\newtheorem{ass}[]{Assumption}
\newtheorem{prpstn}[]{Proposition}
\newtheorem{rmrk}[]{Remark}

\newcommand {\x} { {\bf x} }

\newcommand{\beq}{\begin{equation}}
\newcommand{\eeq}{\end{equation}}

\newcommand\smallO{
  \mathchoice
    {{\scriptstyle\mathcal{O}}}
    {{\scriptstyle\mathcal{O}}}
    {{\scriptscriptstyle\mathcal{O}}}
    {\scalebox{.7}{$\scriptscriptstyle\mathcal{O}$}}
  }

\def\n{{\bf n}}
\def\v{{\bf v}}

\def\red#1{\textcolor{black}{#1}}

\newcommand {\e}  {\varepsilon}

\numberwithin{equation}{section}
\numberwithin{prpstn}{section}
\numberwithin{ass}{section}
\numberwithin{rmrk}{section}

\title{Derivation and application of effective interface conditions for continuum mechanical models of cell invasion through thin membranes \thanks{This work was partially supported by the MIUR grant ``Dipartimenti di Eccellenza 2018-2022'', by the INDAM grant ``Progetto Giovani GNFM 2018'' and by the EPSRC grant no. EP/N014642/1.}}

\author{Mark A. J. Chaplain\thanks{University of St Andrews, UK (majc@st-andrews.ac.uk)}
\and
Chiara Giverso\thanks{Politecnico di Torino, IT (chiara.giverso@polito.it)}
\and
Tommaso Lorenzi\thanks{University of St Andrews, UK (tl47@st-andrews.ac.uk)}
\and
Luigi Preziosi\thanks{Politecnico di Torino, IT (luigi.preziosi@polito.it)}
}

\begin{document}
\date{}
\maketitle

\begin{abstract}
We consider a continuum mechanical model of cell invasion through thin membranes. The model consists of a transmission problem for cell volume fraction complemented with continuity of stresses and mass flux across the surfaces of the membranes. We reduce the original problem to a limiting transmission problem whereby each thin membrane is replaced by an effective interface, and we develop a formal asymptotic method that enables the derivation of a set of biophysically consistent transmission conditions to close the limiting problem. The formal results obtained are validated via numerical simulations showing that the relative error between the solutions to the original transmission problem and the solutions to the limiting problem vanishes when the thickness of the membranes tends to zero. In order to show potential applications of our effective interface conditions, we employ the limiting transmission problem to model cancer cell invasion through the basement membrane and the metastatic spread of ovarian carcinoma. 
\end{abstract}



\section{Introduction}
\paragraph{Biological background} Cell migration is crucial to maintain normal homeostasis~\cite{Li}  and sustain many physiological and pathological 
processes~\cite{Friedl_2012, Kelley, Micalizzi_2010, Nakaya_2013}. During migration phenomena, cells encounter a variety of barriers encompassing other cells, cell-cell junctions, and extracellular matrices (ECMs) of different densities and compositions~\cite{Kelley}. 

One of the most difficult barriers for the cells to cross is the basement membrane. This is a thin, dense and highly cross-linked sheet-like network of ECM macromolecules that underlies, among others, all epithelial and endothelial layers~\cite{Kalluri, Kelley}. With its pore size being in the order of 50 nm, only small molecules such as nutrients (\emph{e.g.} oxygen and glucose) and other chemical factors are able to passively diffuse across the basement membrane \cite{Kalluri,rowe2008breaching}. Nonetheless, such a structural barrier is crossed daily by billions of cells in healthy tissues in the course of normal immune cell trafficking~\cite{Huber}, epithelial-to-mesenchymal transition~\cite{Venkov}, collective cell migration~\cite{Friedl_2012, Micalizzi_2010, Nakaya_2013}, and tissue development and morphogenesis~\cite{Viebahn}. Recent empirical studies~\cite{Kelley, wolf2013physical} suggest that during these physiological processes cells can invade the basement membrane and other thin ECM barriers in a variety of ways, including either active removal (\emph{e.g.} through invadopodia breaching and barrier disruption mediated by the down-regulation of adhesion receptors) or structural remodelling leading to the creation of gaps in the barrier, or even physiological enlargement of preexisting openings that facilitate, for instance, leukocyte trafficking in the vasculature~\cite{Nourshargh}. 

Similar mechanisms of cell invasion are likely to be activated in pathological conditions, including fibrotic diseases (most commonly affecting the lungs or kidneys), inflammatory diseases, arteriosclerosis and neoplastic processes~\cite{Kelley}. In particular, many types of tumours originate and develop in body regions that are separated from the surrounding environment by the basement membrane. This is, for instance, the case of breast tumours (ductal carcinoma)~\cite{Cichon}, ovary tumours~\cite{ahmed2007epithelial}, and exocrine or endocrine pancreatic tumours~\cite{Burns}. During the first stages of cancer progression, non-invasive dysplastic cells proliferate locally and form a carcinoma {\it in situ}. At some later stage of tumour development, such a localised cancer lesion may acquire the capacity to invade the adjacent tissues by perforating the basement membrane, thus becoming an invasive carcinoma~\cite{Christofori, Spaderna}. The transition from carcinoma {\it in situ} to invasive carcinoma is sustained by the ability of cancer cells to produce matrix metalloproteinases (MMPs). These are enzymes capable of digesting the collagen fibers that constitute the extracellular environment and the basement membrane~\cite{Ilina_2011, wolf2013physical}. The MMPs' action widens the pores of the fibre networks and enable cancer cells to spread from the primary site to the surrounding tissues. Notably, experimental studies on cancer cell mobility in MMP-degradable collagen lattices and non-degradable substrates of various porosity have revealed the existence of an ECM critical pore size below which cancer cell migration is entirely hampered in the absence of MMP secretion. Such a critical pore size was termed ``the physical limit of migration''~\cite{wolf2013physical}.

\smallskip
\paragraph{Mathematical modelling background} Despite our growing knowledge about the underpinnings of cell invasion during physiological and pathological processes~\cite{Friedl_2012, Hagedorn_2011, Kelley, rowe2008breaching, wolf2013physical, Wolf_2007}, a number of key aspects still remain unclear. This is mainly due to the difficulty of examining {\it in vivo} the interactions occurring between cells and the basement membrane or other ECM barriers during cellular invasion, as well as to the wide range of diverse mechanisms that cells can use to cross different extracellular structures~\cite{Kelley}. As a consequence of our partial understanding of this complex biological phenomenon, there has been little prior work on the mathematical modelling of cell invasion through thin membranes. In fact, classical mathematical models of tumour growth~\cite{Araujo, Franks, Norton, CGP06} and cell migration on two-dimensional flat substrates~\cite{Danuser} do not take into account the effect of cell invasion through ECM barriers nor the transition from carcinomas {\it in situ} to invasive tumours. 

Only more recently physiological and pathological processes involving the migration of single cells in the presence of obstacles or barriers have been mathematically described by means of discrete models~\cite{giverso2014influence, jagiella2016inferring, Gomez}, and different aspects of tumour growth in confined environments have been investigated {\it in silico} using both discrete and hybrid models~\cite{giverso2010individual, Kim_Othmer_2013, Kim_Othmer_2011}. These models can be easily tailored to capture fine details of the changes in cell-cell and cell-ECM adhesion properties observed during cell migration. However, their computational cost can become prohibitive for large cell numbers. Therefore, to model cell migration through the basement membrane and other thin ECM barriers at the scale of larger portions of tissues, it is desirable to use continuum models, which offer the possibility to carry out efficient numerical simulations for large cell numbers that are biologically and clinically relevant. 

In this regard, focussing on breast cancer, which originates in the epithelial lining of the milk ducts, Ribba \emph{et al.}~\cite{Ribba2006} have proposed a mathematical model of cancer cell invasion whereby the basement membrane of the ducts is explicitly represented as a weakly permeable thin region. Although it has provided some interesting biological insights, such a modelling approach could become computationally inefficient in the presence of multiple thin membranes, as they would still be modelled as finite regions. Moreover, Gallinato \emph{et al.}~\cite{Gallinato} have proposed a mixture model of breast cancer cell invasion whereby the presence of the basement membrane of the milk ducts is taken into account by imposing nonlinear Kedem-Katchalsky interface conditions~\cite{cangiani2010spatial, dimitrio2013spatial, eliavs2014dynamics, kedem1958thermodynamic, Kleinhans, Perrussel} at the interface between the tumour and the host region. In the setting of Gallinato \emph{et al.}~\cite{Gallinato}, such transmission conditions lead the normal velocity of the cells and the cell volume fraction to be continuous across the basement membrane, which is not necessarily the case. Finally, Arduino \& Preziosi~\cite{arduino2015multiphase} and Giverso \emph{et al.}~\cite{giverso2018nucleus} have presented a number of multiphase models of cancer cell migration and invasion through the ECM. In agreement with the biological experiments of Wolf \emph{et al.} \cite{wolf2013physical}, in these models the cellular mobility vanishes when the ECM pore size decreases below a certain critical value. These models effectively capture the fact that the ECM critical pore size is relative to the geometrical and mechanical characteristics of the cells (\emph{e.g.} the size and elasticity of the nucleus, the stiffness of the nuclear membrane, cellular adhesion and traction), and they have been proven useful to study cancer cell invasion in cases where the morphological characteristics of the ECM are spatially heterogeneous, or even discontinuous. However, such models do not apply to biological scenarios where ECM regions with different mechanical and structural properties (\emph{i.e.} different cell mobilities) are separated by thin membranes.

\smallskip
\paragraph{Contents of the paper} In this paper, we consider a continuum mechanical model of cell movement and proliferation in a spatial domain that is divided into subdomains by one or multiple thin membranes. The model is formulated in terms of a transmission problem defined by a system of nonlinear partial differential equations for the cell volume fraction complemented with mass-continuity and stress-continuity conditions on the interfaces between the membranes and the rest of the domain. 

Nonlinear partial differential equations describing reaction-diffusion processes and transport phenomena in spatial domains that comprise different parts separated by thin layers (\emph{i.e.} films or membranes) arise in the mathematical modelling of various chemical, physical and biological systems~\cite{abdelkader2013asymptotic, achdou1998effective, aho2016diffusion, ammari2004reconstruction, bellieud2016asymptotic, Scialo, bonnet2016effective,
bruna2015effective, capdeboscq2006pointwise, caubet2017new, delourme2012approximate, gahn2018effective, geymonat1999mathematical, haddar2008generalized, joly2006matching, lenzi2016anomalous, marigo2016two, neuss2007effective, neuss2010multiscale, Ochoa, perrussel2013asymptotic, poignard2012boundary}. Due to the analytical and numerical challenges posed by the presence of such layers~\cite{auvray2018asymptotic}, it is often convenient to approximate the original problem by an equivalent transmission problem whereby each thin layer is replaced by an effective interface. The equivalent problem is then closed by imposing appropriate transmission conditions on the effective interfaces. 

In this spirit, we develop a formal procedure to derive a set of biophysically consistent interface conditions to close the limiting problem. Specifically, we find that the mass flux across the effective interfaces must be continuous, as one would expect, and proportional to the jump of a term linked to the cell pressure. The biophysical interest lies in the fact that this  proportionality coefficient can be related to the size of the pores of the thin membrane, as well as to the geometrical and mechanical characteristics of the cells as in~\cite{arduino2015multiphase, giverso2018nucleus, giverso2014influence}. This makes the limiting transmission problem suitable for providing a possible macroscopic description of cell invasion through thin membranes that takes explicitly into account cell microscopic characteristics, such as the mechanical constraints imposed by the cell nuclear envelope and the solid material surrounded by it~\cite{wolf2013physical}. 

The transmission condition identified by the limiting procedure can be regarded as a nonlinear generalisation of the classical Kedem-Katchalsky interface condition, as it reduces to it for a peculiar (logarithmic) choice of the constitutive relation between the cell pressure and the cell volume fraction. In contrast to other nonlinear Kedem-Katchalsky interface conditions that have been previously employed to model cell invasion through the basement membrane~\cite{Gallinato}, our transmission condition allows the cell volume fraction to be discontinuous across the equivalent interface, while ensuring mass conservation.

The remainder of the paper is organised as follows. In Section~2, we present the original transmission problem and we introduce the related limiting problem. In Section~3, we formally derive a set of effective interface conditions to close the limiting problem. In Section~4, we present sample numerical solutions that illustrate the formal results established in Section~3 and show their potential applications. In particular, we use the limiting transmission problem to describe cancer cell invasion through the basement membrane and to model the metastatic spread of ovarian carcinoma. Section~5 concludes the paper and provides a brief overview of possible research perspectives.

\section{Statement of the problem}
\begin{figure}[h!]
\centering     
\includegraphics[width=0.4\textwidth]{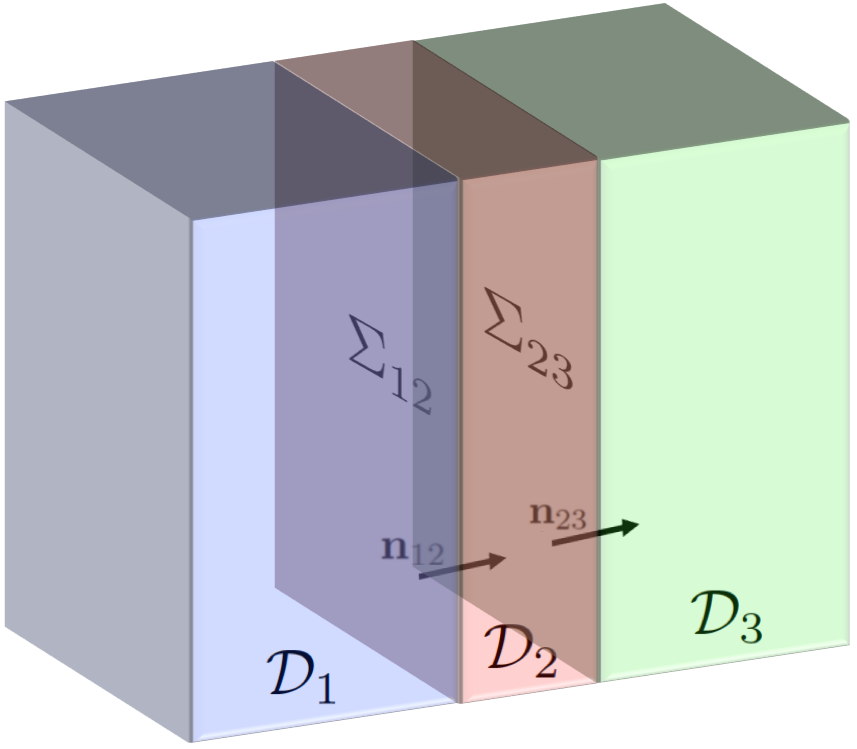}
\caption{{\bf Example of spatial domain and related notation.}}
\label{fig:Domain}
\end{figure}

We consider a population of cells moving through a region of space that is filled with a porous embedding medium, \emph{e.g.} the ECM. Mathematically, we identify such a region with a simply-connected spatial domain ${\cal D} \subset \mathbb{R}^{d}$, with $d=1,2,3$. Focussing on the biological scenario where the spatial domain is divided into two regions separated by a porous membrane, we let the domain ${\cal D}$ consist of three subdomains represented as the open sets ${\cal D}_1$, ${\cal D}_2$ and ${\cal D}_3$, as in the scheme depicted in Fig.~\ref{fig:Domain} for a three-dimensional case. The subdomain ${\cal D}_2$ represents the porous membrane, and the interfaces between the membrane and the subdomains ${\cal D}_1$ and ${\cal D}_3$ are denoted by $\Sigma_{12}$ and $\Sigma_{23}$, respectively. 

We model the cell volume fraction at position $\x \in {\cal D}$ and time $t\geq0$ by means of the function $\rho(t,\x) \geq 0$. The evolution of the cell volume fraction is governed by the mass balance equation  
\begin{equation}
\label{e.CPori}
\frac{\partial \rho}{\partial t} \, + \, \nabla \cdot (\rho \, \v)= \Gamma(\rho), \quad (t,\x) \in \mathbb{R}^+ \times {\cal D}
\end{equation}
complemented with the momentum-related equation \red{for an elastic fluid, neglecting inertia,}
\begin{equation}
\label{def.v}
\v := - \mu \, \nabla p,
\end{equation}
\red{and a barotropic relation $p \equiv p(\rho)$ for the cell pressure $p$.} If necessary, one can let the net growth rate $\Gamma$ depend also on the concentrations of some chemical factors, such as nutrients and growth factors, and couple Eq.~\eqref{e.CPori} with the mass balance equations modelling their evolution. In analogy with the classical Darcy's law for fluids, the function $\mu(t,\x) \geq 0$ is the cell mobility coefficient and Eq.~\eqref{def.v} models the tendency of cells to move towards regions where they feel less compressed~\cite{Ambrosi_closure}. 
\red{
\begin{rmrk}
We remark that Eq.~\eqref{def.v} is only an approximate representation of the far more complex process underlying the migration of cellular aggregates, which is governed by a multitude of sub-cellular pathways involving different proteins and chemical species~\cite{Welf, Vorotnikov} and is influenced by the mechanical properties both of the single cells and of the sub-cellular elements of the aggregate~\cite{giverso2018nucleus}, as well as by the conditions of the surrounding environment. However, when looking at cell migration at the tissue scale, the ensemble of cells that constitute a cellular aggregate can be described as a single phase material -- or possibly a multi-phase material -- with liquid~\cite{Lowengrub2010, Byrne_nonnecrotic, Cristini, Friedman, Greenspan,  Chen2001, Giverso_avascular}, or elastic/hyperelastic~\cite{Ambrosi2004, araujo2004, Humphrey}, or elasto-viscoplastic~\cite{Giverso_necrotic} characteristics. In particular, the use of a liquid-like constitutive assumption is supported by experimental evidence~\cite{Foty, Aigouy2013, Delanoe2013, Joanny_Pnas} indicating that cellular aggregates behave like elastic solids  over short timescales (i.e. time scales of the order of a few minutes) but eventually display a fluid-like behaviour (i.e. over time scales of the order of cell division and apoptosis). For this reason, the representation of living materials as viscous/inviscid/elastic fluids is now commonly employed~\cite{Lowengrub2010}. 
\end{rmrk}
}
It is important to stress the fact that we let the cell mobility coefficient be a function of both $t$ and $\x$. This is to take into account the heterogeneous composition of the spatial domain ${\cal D}$ and the biological notion that the mobility of cells in the embedding medium, especially within the membrane, can vary considerably across space and time. Variability of the cell mobility can be due both to local variations in the micro-structure of the ECM and to spatio-temporal changes in the concentration of MMPs. Therefore, one may let the function $\mu$ depend explicitly on the local concentration of MMPs and then couple Eqs.~\eqref{e.CPori} and \eqref{def.v} with a conservation equation for the MMP concentration, as we will do in Section 4. 

From continuum mechanics, one has that mass flux \red{and stresses} must be continuous across the interfaces $\Sigma_{12}$ and $\Sigma_{23}$. Within the framework of Eqs.~\eqref{e.CPori} and \eqref{def.v}, such continuity conditions translate into the following interface conditions:
\begin{equation}\label{flux_continuity}
[\![\rho \, \v\cdot\n_{ij}]\!]=0 \; \mbox{ on } \Sigma_{ij} \; \text{ with } \; i=1,2 \; \text{ and } \; j=i+1,
\end{equation}
and
\begin{equation}\label{stress_continuity}
[\![p]\!]=0 \; \mbox{ on } \Sigma_{ij} \; \text{ with } \; i=1,2 \; \text{ and } \; j=i+1.
\end{equation}
In Eqs.~\eqref{flux_continuity} and \eqref{stress_continuity}, the notation $[\![(\cdot)]\!]$ represents the jump across the interface $\Sigma_{ij}$, \emph{i.e.} $[\![(\cdot)]\!] := (\cdot)_{j} - (\cdot)_{i}$,
with the subscript $i$ indicating that $(\cdot)$ is evaluated as the limit to a point of the interface coming from the subdomain ${\cal D}_i$. Moreover, as shown in Fig.~\ref{fig:Domain}, we denote by $\n_{ij}$ the unit vector normal to the interface $\Sigma_{ij}$ that points towards the subdomain ${\cal D}_{j}$. Substituting the expression~\eqref{def.v} for the velocity field $\v$ into the flux-continuity condition~\eqref{flux_continuity} yields 
\begin{equation}\label{flux}
 [\![\mu \, \rho \, \nabla p \cdot\n_{ij}]\!]=0\; \mbox{ on } \Sigma_{ij} \; \text{ with } \; i=1,2 \; \text{ and } \; j=i+1.
\end{equation}
In order to close the transmission problem defined by Eqs.~\eqref{e.CPori} and \eqref{def.v} complemented with the interface conditions~\eqref{stress_continuity} and \eqref{flux}, in addition to prescribing suitable boundary conditions on the outer boundaries (\emph{i.e.} the non-interfacing parts of the boundaries of the three spatial subdomains) and suitable initial conditions, one should specify a barotropic relation $p(\rho)$.

In general, the three subdomains can differ in their biophysical properties. As a result, the mobility coefficient and the net growth rate can become discontinuous across the interfaces $\Sigma_{12}$ and $\Sigma_{23}$. In this case, denoting by $\rho_i(t,\x)$, $\mu_i(t,\x)$ and $\Gamma_i(\rho_i)$ the restrictions to the subdomain ${\cal D}_i$ of the functions that represent the local cell volume fraction, the mobility coefficient and the net growth rate, respectively, we can rewrite the problem defined by Eqs.~\eqref{e.CPori} and \eqref{def.v} subject to the interface conditions~\eqref{stress_continuity} and \eqref{flux} as 
\begin{equation}\label{tlp}
\left\{
\begin{array}{ll}
\displaystyle{\frac{\partial \rho_i}{\partial t} - \nabla \cdot(\mu_i \, \rho_i \, \nabla p) = \Gamma_i(\rho_i)} 
&{\rm in}\ {\cal D}_i, \; i=1,2,3,\\[12pt]
\mu_{i} \, \rho_{i} \, \nabla p \cdot {\bf n}_{ij} = \mu_{j} \, \rho_{j} \, \nabla p \cdot {\bf n}_{ij} 
&{\rm on}\ \Sigma_{ij}, \; i=1,2,\\[12pt]
[\![p]\!] = 0&{\rm on}\ \Sigma_{ij}, \; i=1,2,
\end{array}
\right.
\end{equation}
with $j=i+1$. We make the following assumptions:
\begin{ass}
\label{ass:mu}
The cell mobility coefficient $\mu_i$ is continuous in both arguments for all $i=1,2,3$.
\end{ass}
\begin{ass}
\label{ass:Gammai}
The net growth rate $\Gamma_i$ is a continuously differentiable function of the cell volume fraction for all $i=1,2,3$.
\end{ass}
\begin{ass}
\label{ass.p}
\red{The pressure $p$ is given by a barotropic relation $p \equiv f(\rho)$ where $f$ is a continuously differentiable and monotonically increasing function of the cell volume fraction.}
\end{ass}

\begin{rmrk}
In the case where the pressure $p$ is a continuous function of the cell volume fraction $\rho$, the stress-continuity condition~\eqref{stress_continuity} implies that also $\rho$ is continuous across the interfaces $\Sigma_{12}$ and $\Sigma_{23}$, that is, 
$$
[\![\rho]\!]=0  \; \mbox{ on } \Sigma_{ij} \; \text{ with } \; i=1,2 \; \text{ and } \; j=i+1.
$$
Hence, the flux-continuity conditions~\eqref{flux_continuity} or \eqref{flux} read as
$$
[\![\v\cdot\n_{ij}]\!]=0 \quad \text{or} \quad [\![\mu \, \nabla p \cdot\n_{ij}]\!]=0  \; \mbox{ on } \Sigma_{ij} \; \text{ with } \; i=1,2 \; \text{ and } \; j=i+1.
$$
\end{rmrk}

In most biologically relevant scenarios arising in the study of cell invasion through the basement membrane and other ECM barriers, the thickness of the membrane or the barrier is much smaller than the characteristic size $L > 0$ of the spatial domain. In order to translate this biological observation into mathematical terms, we define the thickness of the membrane represented as the subdomain ${\cal D}_2$ as
\begin{equation}\label{thickness}
\e := \max_{\hat{\x}_{12} \in \Sigma_{12}} \big\{ \min \{a > 0: \hat{\x}_{12} + a \, {\bf n}_{12} \in \Sigma_{23}\} \big\}
\end{equation}
and we assume $\e \ll L$. In the biological scenarios corresponding to the assumption $\e \ll L$, one typically wishes to:
\begin{itemize}
\item[i)] replace the subdomain ${\cal D}_{2}$ with an effective interface, which is obtained from the actual interfaces $\Sigma_{12}$ and $\Sigma_{23}$ by letting $\e \to 0$;
\item[ii)] find biophysically consistent transmission conditions to impose on the effective interface in this asymptotic regime.
\end{itemize}

With these goals in mind, we rewrite the transmission problem~\eqref{tlp} as
\begin{equation}\label{tlpeps}
{\cal P}_\e\equiv\left\{
\begin{array}{ll}
\displaystyle{\frac{\partial \rho_{i \e}}{\partial t} - \nabla \cdot(\mu_{i \e} \rho_{i \e} f'(\rho_{i\e}) \nabla \rho_{i \e}) = \Gamma_{i \e}(\rho_{i \e})} 
&{\rm in}\ {\cal D}_{i \e}, \; i=1,2,3,\\[12pt]
\mu_{i \e} \nabla \rho_{i \e} \cdot {\bf n}_{ij} =\mu_{j \e} \nabla \rho_{j \e} \cdot {\bf n}_{ij} &{\rm on}\ \Sigma_{ij \e}, \; i=1,2,\\[12pt]
\rho_{i \e} = \rho_{j \e} &{\rm on}\ \Sigma_{ij \e}, \; i=1,2,
\end{array}
\right.
\end{equation}
with $j=i+1$, while the limiting transmission problem whereby the subdomain ${\cal D}_{2 \e}$ is replaced by an effective interface reads as
\begin{equation}\label{thinlpo}
{\cal P}_0\equiv\left\{
\begin{array}{ll}
\displaystyle{\frac{\partial \tilde \rho_i}{\partial t} - \nabla \cdot(\tilde \mu_i \, \tilde \rho_i \, f'(\tilde \rho_{i}) \nabla \tilde \rho_{i}) = \tilde \Gamma_i(\tilde \rho_{i})} 
&{\rm in}\ \tilde {\cal D}_i, \; i=1,3,\\[12pt]
\text{transmission conditions} &{\rm on}\ \tilde \Sigma_{13},
\end{array}
\right.
\end{equation}
where
\begin{equation}\label{limits1}
\tilde {\cal D}_1 = \lim_{\e \to 0} {\cal D}_{1 \e}, \quad \tilde {\cal D}_3 = \lim_{\e \to 0} {\cal D}_{3 \e}, \quad \tilde \Sigma_{13} = \lim_{\e \to 0} \Sigma_{12 \e} = \lim_{\e \to 0} \Sigma_{23 \e},
\end{equation}
\begin{equation}\label{limits0}
\tilde \rho_i = \lim_{\e \to 0} {\rho}_{i \e}, \quad \tilde {\mu}_i = \lim_{\e \to 0} {\mu}_{i \e} \quad \text{and} \quad \tilde {\Gamma}_i(\tilde \rho_i) = \lim_{\e \to 0} {\Gamma}_{i \e}(\rho_{i \e}).
\end{equation}

\begin{rmrk}
In the remainder of the paper, we will refer to the transmission problem ${\cal P}_\e$ defined by~\eqref{tlpeps}, or equivalently by~\eqref{tlp}, as the ``thin layer problem'', and to the limiting transmission problem ${\cal P}_0$ defined by~\eqref{thinlpo} along with the appropriate transmission conditions as the ``effective interface problem''.
\end{rmrk}

The next section will be devoted to derive the transmission conditions that are necessary to complete the effective interface problem ${\cal P}_0$. \red{For the effective interface $\tilde \Sigma_{13}$ (\emph{i.e.} an infinitesimal region) to have an effect on cell invasion analogous to that of the actual thin membrane represented as the subdomain ${\cal D}_{2 \e}$ (\emph{i.e.} a finite region), when letting $\e \to 0$ we will need to compact the membrane (\emph{vid.} Remark~\ref{remeffperm}). In other words, we will obtain the effective interface by virtually shrinking the pores of the membrane in such a way as to cause a reduction in the local permeability $\mu_{2 \e}$ that is proportional to the local shrinkage. This ensures that the existing relationships between the structural characteristics of the thin membrane and the biophysical properties of the cells will remain intact across $\tilde \Sigma_{13}$. To this end, we will assume}
\red{
\begin{equation}
\label{eq:assTC2}
\mu_{2 \e} \xrightarrow[\e \to 0]{} 0 \; \text{ in such a way that } \; \frac{\mu_{2 \e}}{\e} \xrightarrow[\e \to 0]{} \tilde \mu_{13}, \; \mbox{ with } \; \tilde \mu_{13} : \mathbb{R}^+ \times {\cal D}_{2 \e} \to \mathbb{R}^+
\end{equation}
and
\begin{equation}
\label{eq:assTC2b}
\lim_{\e \to 0} \frac{\nabla \mu_{2 \e}}{\e} \cdot {\n}_{12}  = \lim_{\e \to 0} \frac{\nabla \mu_{2 \e}}{\e} \cdot {\n}_{23} = \nabla \tilde \mu_{13} \cdot {\tilde \n}_{13} = 0,
\end{equation}
where $\tilde{\n}_{13}$ is the unit vector normal to the interface $\tilde \Sigma_{13}$ that points towards the subdomain $\tilde {\cal D}_{3}$. 
}
The positive bounded function $\tilde \mu_{13}$ can be seen as the ``effective mobility coefficient'' of the cells through the thin membrane represented as the effective interface $\tilde \Sigma_{13}$. 

\begin{rmrk}
\label{remeffperm}
By analogy, consider a liquid flowing through a layer of porous material with unitary cross-sectional area. The liquid flux $Q$ can be computed using the classical Darcy's law as
$$
Q=-\,\dfrac{\kappa}{\nu} \dfrac{\Delta P}{\Delta x} \,,
$$
where $\Delta P$ is the pressure drop between the ends of the layer, $\Delta x$ is the thickness of the layer, $\kappa$ represents the hydraulic permeability of the material and $\nu$ is the dynamic viscosity of the liquid. We can draw a conceptual analogy between the biological problem at hand and the case of the liquid by noting that \red{in order to preserve the flux $Q$ when taking the limit $\Delta x \to 0$ the key is to keep the pressure drop $\Delta P$ fixed. This can be achieved by letting 
$$
\kappa \xrightarrow[\Delta x \to 0]{} 0 \; \text{ in such a way that } \; \frac{\kappa}{\Delta x} \xrightarrow[\Delta x \to 0]{} \tilde \kappa, \; \mbox{ with } \; \tilde \kappa \in \mathbb{R}^+,
$$
where $\tilde \kappa$ represents an ``effective permeability'' of the porous layer in the case where the layer is thin. The latter assumption is analogous to assumption~\eqref{eq:assTC2}.}
\end{rmrk}

\section{Formal derivation of the interface conditions for the effective transmission problem}
\label{dercond}
In this section, we formally derive the transmission conditions required to complete the effective transmission problem ${\cal P}_0$ defined by~\eqref{thinlpo}. \red{In summary, as established by Proposition~\ref{Th1}, we show that the mass flux across the effective interface $\tilde \Sigma_{13}$ is continuous and we find an additional transmission condition that establishes a relationship between the mass flux across the effective interface $\tilde \Sigma_{13}$ and the effective cell mobility coefficient $\tilde \mu_{13}(t,\x)$. }
\red{
\begin{prpstn}\label{Th1} 
Under Assumptions \ref{ass:mu}-\ref{ass.p}, the following transmission condition formally applies to the effective interface problem~\eqref{thinlpo}
\begin{equation}
\label{eq:TC1}
\tilde \mu_1 \, \tilde \rho_1 \, f'(\tilde \rho_1) \, \nabla \tilde \rho_1 \cdot \tilde{\n}_{13}= \tilde \mu_3 \, \tilde \rho_3 \, f'(\tilde \rho_3) \, \nabla \tilde \rho_3 \cdot\tilde{\n}_{13} \qquad  \text{on } \tilde \Sigma_{13}.
\end{equation}
Moreover, under the additional assumptions~\eqref{eq:assTC2} and~\eqref{eq:assTC2b}, 
\begin{equation}
\label{eq:TC2}
\tilde \mu_{13} \, [\![\Pi]\!] = \tilde \mu_1 \, \tilde \rho_1 \, f'(\tilde \rho_1) \, \nabla \tilde \rho_1 \cdot\tilde{\n}_{13} = \tilde \mu_3 \, \tilde \rho_3 \, f'(\tilde \rho_3) \, \nabla \tilde \rho_3 \cdot\tilde{\n}_{13} \qquad  \text{on } \tilde \Sigma_{13},
\end{equation}
where the function $\Pi(\rho)$ is defined according to the equation
\begin{equation}\label{e.Pip}
\Pi'(\rho)  := \rho \, f'(\rho).
\end{equation}
\end{prpstn}
}
\red{
\begin{proof}
For ease of presentation, we formally derive the interface conditions~\eqref{eq:TC1} and~\eqref{eq:TC2} in the case where $\Sigma_{12 \e}$ and $\Sigma_{23 \e}$ are parallel planes, but there would be no additional difficulty in considering more general cases. We introduce the notation ${\cal D}_{2 \e} \ni \x := (x_\perp, \mathbf{x}_{\Sigma})$, where $x_\perp := \x \cdot \n_{12} = \x \cdot \n_{23}$. We also make the change of variables $x_{\perp} \mapsto x_{\perp} - \hat{x}_{12 \perp}$, with $\hat x_{12 \perp}$ given by $\hat{\x}_{12} = (\hat x_{12 \perp}, \hat \x_{12 \Sigma}) \in \Sigma_{12 \e}$, and let $\eta := \dfrac{x_{\perp}}{\e}  \in (0,1)$, so that Eq.~\eqref{tlpeps} for $\rho_{2 \e}$ can be rewritten as
\beq
\label{res1}
\displaystyle{\frac{\partial \rho_{2 \e}}{\partial t} - \nabla_{\x_\Sigma} \cdot (\mu_{2 \e} \rho_{2 \e} f'(\rho_{2\e}) \nabla_{\x_\Sigma} \rho_{2 \e}) - \frac{1}{\e}\frac{\partial}{\partial \eta} \left(\frac{\mu_{2 \e}}{\e} \rho_{2 \e} f'(\rho_{2\e}) \frac{\partial  \rho_{2 \e}}{\partial \eta}\right)= \Gamma_{2 \e}(\rho_{2 \e})} 
\eeq
and the related flux continuity conditions can be rewritten as
\beq
\label{res2}
\displaystyle{\dfrac{\mu_{2\e}}{\e} \rho_{2\e} f'(\rho_{2\e} ) \frac{\partial \rho_{2 \e}}{\partial \eta}\Big|_{\eta=0} = \mu_{1\e}  \rho_{1\e} f'(\rho_{1\e}) \nabla \rho_{1\e} \cdot \mathbf{n}_{12} \big|_{\Sigma_{12 \e}} \, }, 
\eeq
\beq
\label{res3}
\displaystyle{\dfrac{\mu_{2\e}}{\e} \rho_{2\e} f'(\rho_{2\e} ) \frac{\partial \rho_{2 \e}}{\partial \eta}\Big|_{\eta=1} = \mu_{3\e}  \rho_{3\e} f'(\rho_{3\e}) \nabla \rho_{3\e} \cdot \mathbf{n}_{23} \big|_{\Sigma_{23 \e}} \, }.
\eeq
Rearranging terms in~\eqref{res1} yields
\beq
\label{res4}
\frac{\partial}{\partial \eta} \left(\frac{\mu_{2 \e}}{\e} \rho_{2 \e} f'(\rho_{2\e}) \frac{\partial  \rho_{2 \e}}{\partial \eta}\right) = \e \left(\frac{\partial \rho_{2 \e}}{\partial t} - \nabla_{\x_\Sigma} \cdot (\mu_{2 \e} \rho_{2 \e} f'(\rho_{2\e}) \nabla_{\x_\Sigma} \rho_{2 \e}) - \Gamma_{2 \e}(\rho_{2 \e})\right).
\eeq
We make the ansatz
\beq
\label{res5}
\rho_{2\varepsilon}\left(\dfrac{x_{\perp}}{\e}, \mathbf{x}_{\Sigma}\right) = \rho_2^0(\eta, \mathbf{x}_{\Sigma}) + \varepsilon \rho_2^1(\eta, \mathbf{x}_{\Sigma}) + \smallO(\varepsilon)
\eeq
and compute the asymptotic expansions 
\beq
\label{res6}
f'(\rho_{2\e}) = f'(\rho_2^0)+ \varepsilon f''(\rho_2^0) \rho_2^1 + \smallO(\varepsilon), \qquad \Gamma(\rho_{2\e}) = \Gamma(\rho_2^0)+ \varepsilon \Gamma'(\rho_2^0) \rho_2^1 + \smallO(\varepsilon).
\eeq
Substituting~\eqref{res5} and~\eqref{res6} into~\eqref{res4}, and letting $\e \to 0$, under assumption~\eqref{eq:assTC2} we formally obtain 
\beq
\label{res7}
 \frac{\partial}{\partial \eta} \left(\tilde \mu_{13} \, \rho^0_{2} \, f'(\rho^0_{2}) \, \frac{\partial  \rho^0_{2}}{\partial \eta}\right) = 0 \quad \Longrightarrow \quad \displaystyle{\tilde \mu_{13} \, \rho^0_{2} \, f'(\rho^0_{2}) \, \frac{\partial  \rho^0_{2}}{\partial \eta}} = const. \;\; \forall \eta \in (0,1).
\eeq
In a similar way, from the flux continuity conditions~\eqref{res2} and~\eqref{res3} we formally obtain
\beq
\label{res8}
\displaystyle{\tilde \mu_{13} \, \rho^0_{2} \, f'(\rho^0_{2}) \, \frac{\partial  \rho^0_{2}}{\partial \eta} \Big|_{\eta=0} = \tilde \mu_{1}  \tilde \rho_{1} f'(\tilde \rho_{1}) \nabla \tilde \rho_{1} \cdot \tilde{\mathbf{n}}_{13} \big|_{\tilde \Sigma_{13}} \, },
\eeq
\beq
\label{res9}
\displaystyle{\tilde \mu_{13} \, \rho^0_{2} \, f'(\rho^0_{2}) \, \frac{\partial  \rho^0_{2}}{\partial \eta} \Big|_{\eta=1} = \tilde \mu_{3}  \tilde \rho_{3} f'(\tilde \rho_{3}) \nabla \tilde \rho_{3} \cdot \tilde{\mathbf{n}}_{13} \big|_{\tilde \Sigma_{13}} \, }.
\eeq
Using~\eqref{res7} along with~\eqref{res8} and~\eqref{res9} we find that for all $\eta \in (0,1)$ we have
\beq
\label{res10}
\tilde \mu_{13} \, \rho^0_{2} \, f'(\rho^0_{2}) \, \frac{\partial  \rho^0_{2}}{\partial \eta} = \tilde \mu_1 \, \tilde \rho_1 \, f'(\tilde \rho_1) \, \nabla \tilde \rho_1 \cdot \tilde{\n}_{13} \big|_{\tilde \Sigma_{13}} = \tilde \mu_3 \, \tilde \rho_3 \, f'(\tilde \rho_3) \, \nabla \tilde \rho_3 \cdot \tilde{\n}_{13} \big|_{\tilde \Sigma_{13}}.
\eeq
Hence, the transmission condition~\eqref{eq:TC1} is formally verified. Moreover, under the additional assumption~\eqref{eq:assTC2b}, integrating both sides of~\eqref{res10} with respect to $\eta$ and noting that
$$
\tilde \mu_{13}  \int_0^1 \rho^0_{2} \, f'(\rho^0_{2}) \, \frac{\partial  \rho^0_{2}}{\partial \eta} \, {\rm d}\eta = \tilde \mu_{13}  \int_0^1 \frac{\partial  \Pi}{\partial \eta} \, {\rm d}\eta = \tilde \mu_{13} \, [\![\Pi]\!], 
$$
with $\Pi$ defined according to~\eqref{e.Pip}, we obtain
$$
\tilde \mu_{13} \, [\![\Pi]\!] = \tilde \mu_1 \, \tilde \rho_1 \, f'(\tilde \rho_1) \, \nabla \tilde \rho_1 \cdot\tilde{\n}_{13} = \tilde \mu_3 \, \tilde \rho_3 \, f'(\tilde \rho_3) \, \nabla \tilde \rho_3 \cdot \tilde{\n}_{13}, \quad \text{ on } \; \tilde \Sigma_{13}.
$$
Hence, the transmission condition~\eqref{eq:TC2} is formally verified as well.  
\end{proof}
}
\begin{rmrk}
If the cell pressure is given by the barotropic relation 
$$
p \equiv f(\rho) \quad \text{with} \quad f(\rho) := P\,{\rm ln}\left(\rho/\rho_0\right) \quad \text{and} \quad P>0, \rho_0>0
$$
then Eq.~\eqref{tlpeps} for $\rho_{i \e}$ becomes a nonlinear reaction-diffusion equation with a nonlinearity only in the reaction term, and $\Pi = P \, \rho + C$ with $C \in \mathbb{R}$. In this case, the interface condition~\eqref{eq:TC2} reduces to the classical Kedem-Katchalsky interface condition, i.e. $\tilde \mu_{13} \, (\tilde \rho_3 - \tilde \rho_1)  = \tilde \mu_1 \, \nabla \tilde \rho_1 \cdot \tilde{\n}_{13} = \tilde \mu_3 \, \nabla \tilde \rho_3 \cdot \tilde{\n}_{13}$ on $\tilde \Sigma_{13}$.
\end{rmrk}

\begin{rmrk}
If $\tilde \mu_{13} \equiv 0$ then the thin membrane represented as the effective interface $\tilde\Sigma_{13}$ is impermeable and we recover no-flux boundary conditions on both sides of $\tilde\Sigma_{13}$, i.e. the cells in each subdomain are compartmentalised.
\end{rmrk}
\bigskip

Taken together, the formal results established by Proposition~\ref{Th1} allow us to complete the effective interface problem ${\cal P}_0$ defined by the transmission problem~\eqref{thinlpo} as follows
\begin{equation}\label{P0}
{\cal P}_0\equiv\left\{
\begin{array}{ll}
\displaystyle{\frac{\partial \tilde \rho_1}{\partial t} - \nabla \cdot(\tilde \mu_1 \, \tilde \rho_1 \, f'(\tilde \rho_1) \nabla \tilde \rho_1) = \tilde \Gamma_1(\tilde \rho_1)} 
&{\rm in}\ \tilde {\cal D}_1,\\[12pt]
\displaystyle{\frac{\partial \tilde \rho_3}{\partial t} - \nabla \cdot(\tilde \mu_3 \, \tilde \rho_3 \, f'(\tilde \rho_3) \nabla \tilde \rho_3) = \tilde \Gamma_3(\tilde \rho_3)} 
&{\rm in}\ \tilde {\cal D}_3,\\[12pt]
\tilde \mu_{13} [\![\Pi]\!]  = \tilde \mu_1 \, \tilde \rho_1 \, f'(\tilde \rho_1) \nabla \tilde \rho_1 \cdot\tilde{\n}_{13} = \tilde \mu_3 \, \tilde \rho_3 \, f'(\tilde \rho_3) \nabla \tilde \rho_3 \cdot\tilde{\n}_{13}
&{\rm on}\ \tilde \Sigma_{13}.
\end{array}
\right.
\end{equation}
\red{In order to illustrate these formal results we constructed numerical solutions of a one-dimensional version of the thin layer problem ${\cal P}_\e$ for decreasing values of $\e$, and we compared the numerical solutions obtained with the numerical solutions of the corresponding effective interface problem ${\cal P}_0$. These results are reported in Section~S.1 of the Supplementary Material and show that the relative error between the numerical solutions of the two transmission problems tends linearly to zero as $\e \to 0$.}

\red{
\begin{rmrk}
The results established by Proposition~\ref{Th1} can also be obtained using a control volume approach analogous to that typically used in continuum mechanics (i.e. considering a control volume that cuts across the subdomain ${\cal D}_{2 \e}$).
\end{rmrk}
}

\section{Application of the effective interface conditions} \label{num}
The numerical solutions presented in this section show potential applications of the formal results established by Proposition~\ref{Th1}. \red{In Section~\ref{num2d}, we construct numerical solutions for a two-dimensional model of cancer cell invasion through a basement membrane and the corresponding effective interface problem. The numerical results obtained indicate that the effective interface problem provides a good approximation of the original transmission problem for membranes of sufficiently small thickness. In Section~\ref{secovary}, we construct numerical solutions for an effective interface problem modelling cell invasion dynamics in ovarian carcinoma. The numerical results obtained support the idea that the model can qualitatively reproduce the key steps of the complex process leading to the metastatic spread of ovarian cancer cells.} All numerical simulations are carried out using the finite element software (FEM) COMSOL Multiphysics$^\circledR$, with the parallel sparse direct solver MUMPS.  The method for constructing numerical solutions is based on the backward differentiation formula with an adaptive time-step, and a refined mesh is used in the region about the effective interface. 

\subsection{Numerical simulation of cancer cell invasion through the basement membrane}
\label{num2d}
We compare the numerical solutions of a thin layer problem modelling a two-dimensional cell invasion process with the numerical solutions of the corresponding effective interface problem. We consider a biological scenario whereby cancer cells, which proliferate according to a logistic law with intrinsic growth rate $r >0$, invade a normal tissue composed of healthy cells in homeostatic equilibrium (\emph{i.e.} cells for which proliferation is balanced by natural death) by squeezing through a damaged part of the basement membrane.  Throughout this section, we use the notation $\x = (x/L, y/L)$ to denote the spatial position non-dimensionalised with respect to the thickness $L>0$ of the region represented as the subdomains ${\cal D}_{1 \e}$ and $\tilde{\cal D}_{1}$, and we non-dimensionalise the time variable with respect to the intrinsic growth rate $r$.

We consider the net growth rate
\begin{equation}\label{Gamma}
\Gamma(\varphi,\rho) := \left(1- \rho\right) \rho \, H(\varphi) \, ,
\end{equation}
where $H(\cdot)$ denotes the Heaviside step function and the function $\varphi(t, \x)$ is an auxiliary level set function that tracks the region of space occupied by cancer cells.
Moreover, we use the barotropic relation
\begin{equation}\label{1ddb}
p \equiv f(\rho) \quad \text{with} \quad  f(\rho):= (\rho - \rho_0)_+ \quad  \text{and} \quad 0 < \rho_0 < 1, 
\end{equation}
where $(\cdot)_+$ is the positive part of $(\cdot)$. \red{We remark that we consider a scenario whereby the cell volume fraction at $t=0$ is equal to or greater than $\rho_0$ for all $\x$. Since $\rho_0<1$, under definition~\eqref{Gamma} both the thin layer problem $\mathcal{P}_{\e}$ and the effective interface problem $\mathcal{P}_{0}$ are such that the cell volume fraction will be greater than or equal to $\rho_0$ for all $t \geq 0$. Under this scenario, the barotropic relation~\eqref{1ddb} is such that Assumption~\ref{ass.p} is satisfied.}

\begin{figure}[h!]
\centering     
\subfigure[]{\label{fig:geometria_2Deps}\includegraphics[width=0.35\textwidth]{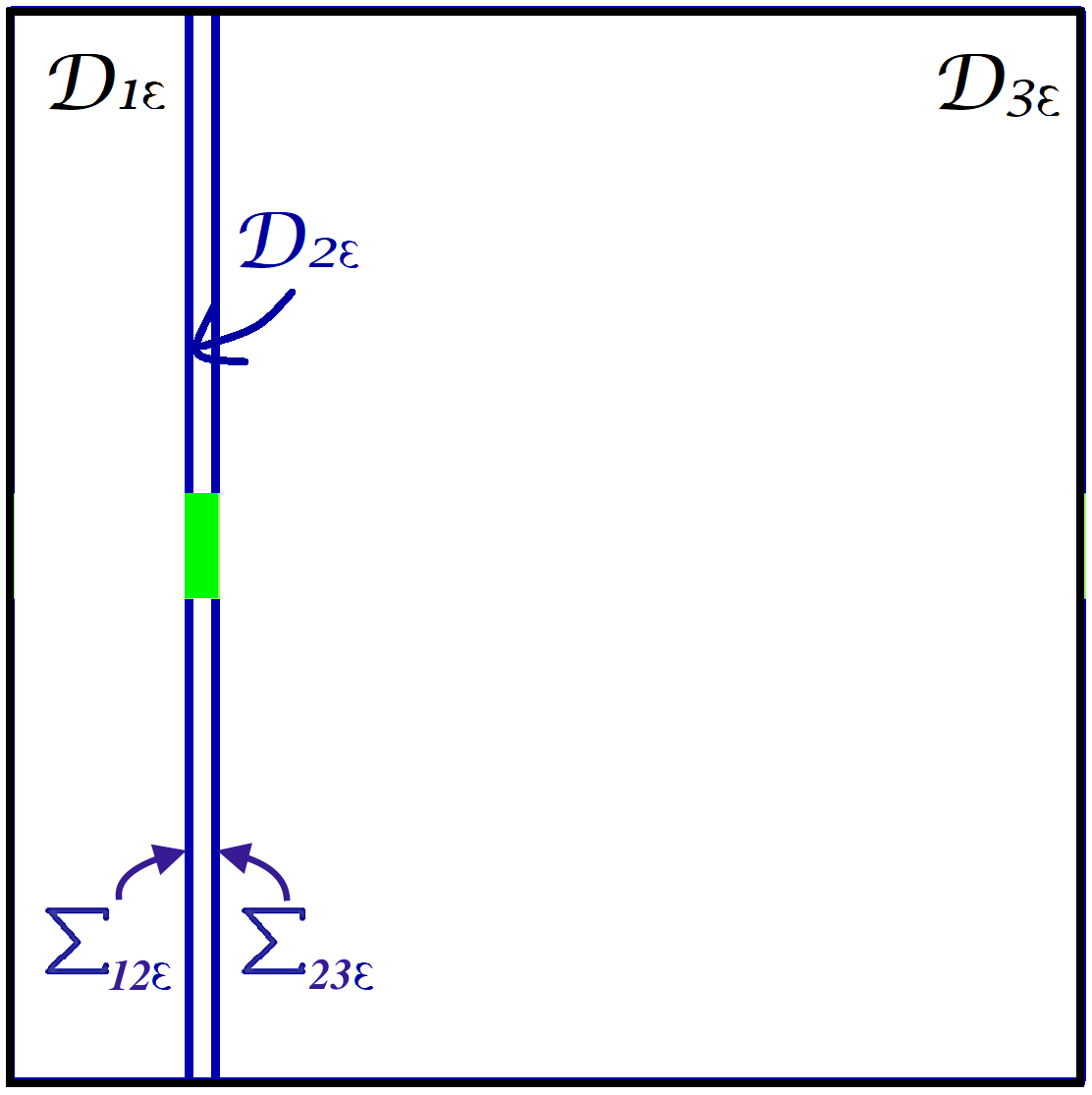}}
\hspace{0.5cm}
\subfigure[]{\label{fig:geometria_2D0}\includegraphics[width=0.35\textwidth]{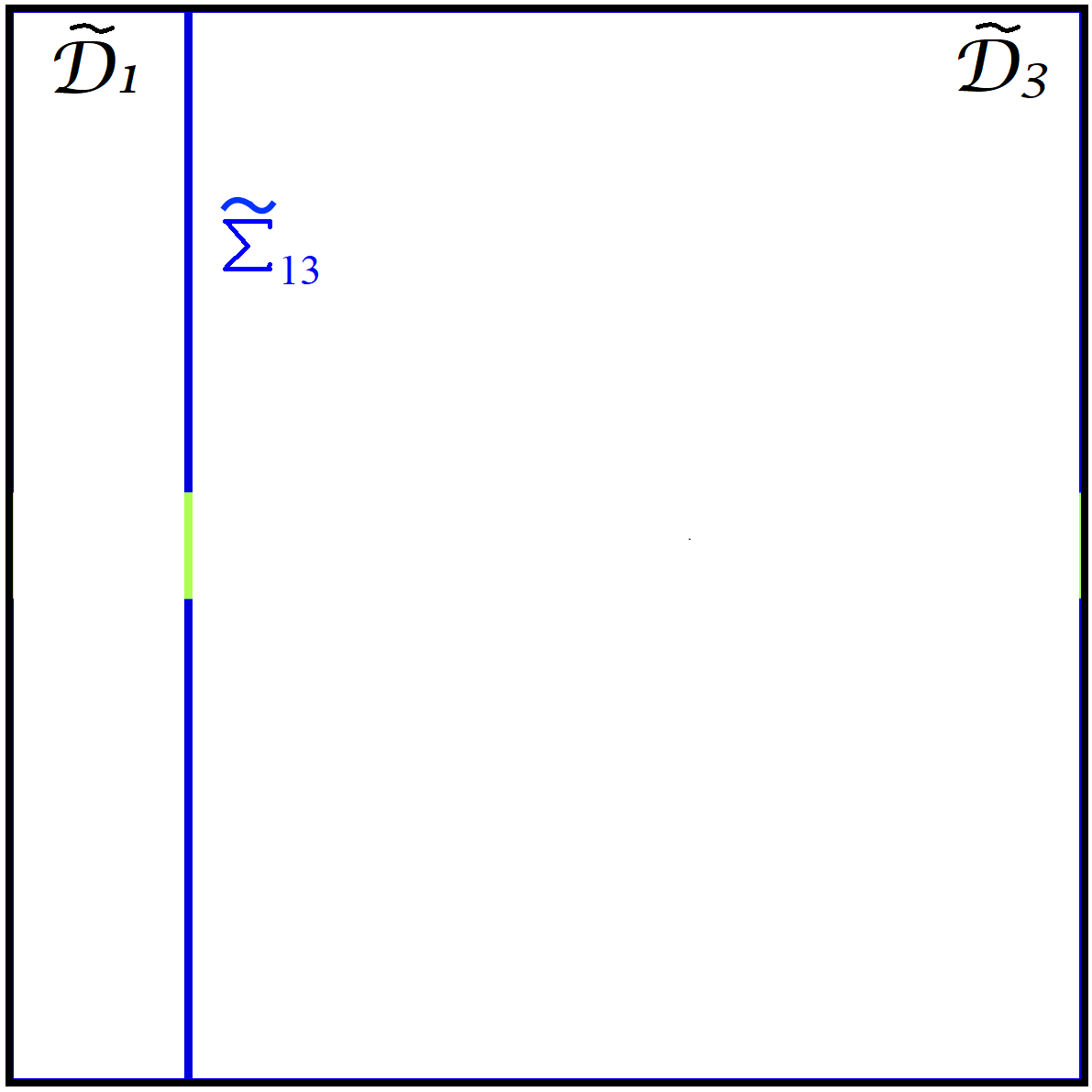}}
\caption{{\bf Spatial domain used in the numerical simulation of cancer cell invasion through the basement membrane.} {\bf (a)} Spatial domain for the thin layer problem. The subdomains ${\cal D}_{1 \e}$ and ${\cal D}_{3 \e}$ are separated by the basement membrane of thickness $\e$, which is represented as the subdomain ${\cal D}_{2 \e}$. The region highlighted in green (i.e. the set $\mathcal{D}_p \subset \overline{{\cal D}}_{2 \e}$) is assumed to be damaged and thus permeable to cancer cells. To construct numerical solutions, we choose ${\cal D}_{1 \e}~:=~(-1,0)~\times (-3,3)$, ${\cal D}_{2 \e}~:=~(0,\e)~\times (-3,3)$, ${\cal D}_3~:=~(\e,5)\times (-3,3)$. {\bf (b)} Spatial domain for the effective interface problem. The subdomains $\tilde{\cal D}_{1}$ and $\tilde{\cal D}_{3}$ are separated by the effective interface $\tilde\Sigma_{13}$. The region highlighted in green (i.e. the set $\tilde\Sigma_{p} \subset \tilde\Sigma_{13}$) is assumed to be damaged and thus permeable to cancer cells. In particular we consider $
\tilde {\cal D}_{1} :=(-1,0) \times (-3,3), \tilde {\cal D}_{3} := (0,5)\times (-3,3)$.\label{fig:geometria_2D}}
\end{figure}

We choose the spatial domains schematised in Fig.~\ref{fig:geometria_2D} to carry out numerical simulations. For the thin layer problem [\emph{vid.} Fig.~\ref{fig:geometria_2Deps}], we let the subdomains ${\cal D}_{1 \e}$ and ${\cal D}_{3 \e}$ be separated by the basement membrane of thickness $\e$, which is represented as the subdomain ${\cal D}_{2 \e}$ with boundaries $\Sigma_{12 \e}$ and $\Sigma_{23 \e}$. We identify the part of the membrane that is damaged, and thus permeable to cancer cells, with a subset $\mathcal{D}_p \subset \overline{{\cal D}}_{2 \e}$. Similarly, for the effective interface problem [\emph{vid.} Fig.~\ref{fig:geometria_2D0}], we let the subdomains $\tilde{\cal D}_{1}$ and $\tilde{\cal D}_{3}$ be separated by the effective interface $\tilde\Sigma_{13}$. In this case, the damaged part of the basement membrane is represented as a set $\tilde\Sigma_{p} \subset \tilde\Sigma_{13}$. 
For simplicity, we assume the cell mobility coefficients in the subdomains ${\cal D}_{1 \e}$ and ${\cal D}_{3 \e}$ to have the same constant value, \emph{i.e.} 
$
\mu_{1 \e}~=~\mu_{3 \e}~\equiv~\bar{\mu} \quad \text{with} \quad \bar{\mu}>0,
$
and we define the mobility coefficient in the subdomain ${\cal D}_{2 \e}$ as
$
\mu_{2 \e}(\x)~:=~\e \, \bar{\mu}_2 \, \mathbf{1}_{\mathcal{D}_p}(\x) \quad \text{with} \quad \bar{\mu}_2 >0
$
where $\mathbf{1}_{\mathcal{D}_p}(\x)$ is a mollification of the indicator function of the set $\mathcal{D}_p \subset \overline{{\cal D}}_{2 \e}$. Accordingly, for the effective interface problem, we choose
$
\tilde \mu_{1}~=~\tilde \mu_{3} \equiv \bar{\mu}, \quad \tilde \mu_{13} := \bar{\mu}_{2} \, \mathbf{1}_{\tilde \Sigma_p}(\x) \, .
$
We assume that cancer cells initially occupy only the region of space on the left of the membrane, while healthy cells reside in the remaining part of the spatial domain. 

We describe the spatio-temporal evolution of the cell volume fraction $\rho_{i \e}(t,\x)$ through the thin layer problem~\eqref{tlpeps} with $f(\rho_{i\e}) $ defined according to~\eqref{1ddb} and $\Gamma_{i \e} \equiv \Gamma(\varphi_\e, \rho_{i \e})$ given by~\eqref{Gamma}.
The function $\varphi_\e(t, \x)$ is the auxiliary level set function that tracks the region of space occupied by cancer cells -- \emph{i.e.} at any time instant $t \geq 0$, if $\varphi_\e(t, \x) > 0$ then the point $\x$ is occupied by cancer cells, whereas if $\varphi_\e(t, \x) \leq 0$ then the point $\x$ is occupied by healthy cells. Hence, the zero level set of the function $\varphi_\e(t, \x)$ corresponds to the boundary of the tumour region at time $t$. The evolution of the function $\varphi_\e(t, \x)$ is governed by the following equation \cite{Osher}
\begin{equation} \label{eq:levelseteps}
\dfrac{\partial \varphi_\e }{\partial t} + \v_{\e}  \cdot \nabla \varphi_\e =0 \; \text{ in } \;  {\cal D}_{1 \e} \cup {\cal D}_{2 \e} \cup {\cal D}_{3 \e}  \quad \text{with } \; \v_{\e}=-\mu_{i\e} f'(\rho_{i\e}) \nabla \rho_{i \e} \; \text{ in } \; {\cal D}_{i \e} 
\end{equation}
for $i=1,2,3$, \red{subject to the continuity conditions
\begin{equation} \label{eq:levelsetepscont}
[\![\varphi_\e]\!] =0  \; \mbox{ on } \Sigma_{ij \e} \; \text{ with } \; i=1,2 \; \text{ and } \; j=i+1.
\end{equation}
}
Notice that the transmission conditions~\eqref{tlpeps}$_2$ ensure the continuity of the normal velocity across the interfaces $\Sigma_{12 \e}$ and $\Sigma_{23 \e}$.

The corresponding effective interface problem is given by the transmission problem~\eqref{P0} with $\tilde \Gamma_i \equiv \Gamma(\tilde{\varphi}, \tilde \rho_i)$ defined according to~\eqref{Gamma} and with $f(\tilde{\rho}_i)$ given by~\eqref{1ddb}. As for the thin layer problem, the function $\tilde \varphi(t, \x)$ is the level set function tracking the region of space occupied by cancer cells, the evolution of which is governed by the following equation \cite{Osher} 
\begin{equation} \label{eq:levelset0}
\dfrac{\partial \tilde \varphi }{\partial t} + \tilde \v  \cdot \nabla \tilde \varphi =0 \; \text{ in } \; \tilde {\cal D}_1 \cup \tilde {\cal D}_3 \quad \text{with } \; \tilde \v= - \tilde \mu_{i} f'(\tilde \rho_i) \nabla \tilde \rho_i \; \text{ in } \; \tilde {\cal D}_{i} \; \text{ for } \; i=1,3,
\end{equation}
\red{subject to the continuity condition
\begin{equation} \label{eq:levelset0cont}
[\![\tilde \varphi]\!] =0 \quad  \mbox{ on } \tilde \Sigma_{13}.
\end{equation}
A formal derivation of condition~\eqref{eq:levelset0cont} is provided in Section~S.2 of the Supplementary Material.} Finally, we choose parameter values, boundary conditions and initial conditions corresponding to those of the thin layer problem.

\begin{figure}[h!]
\centering     
\subfigure[$t=5$]{\label{fig:2Da_eps0p1}\includegraphics[width=0.22\textwidth]{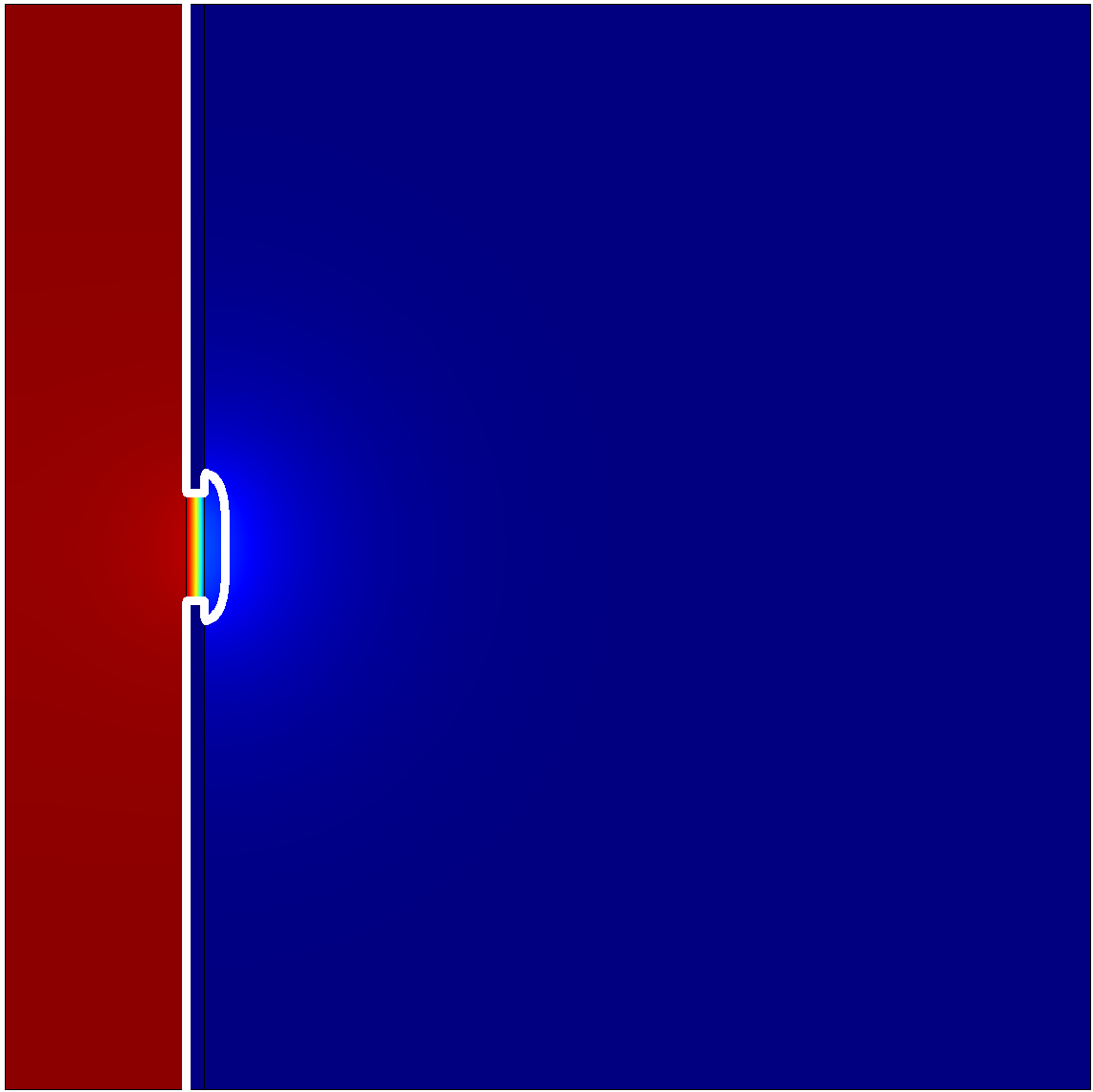}}
\subfigure[$t=10$]{\label{fig:2Db_eps0p1}\includegraphics[width=0.22\textwidth]{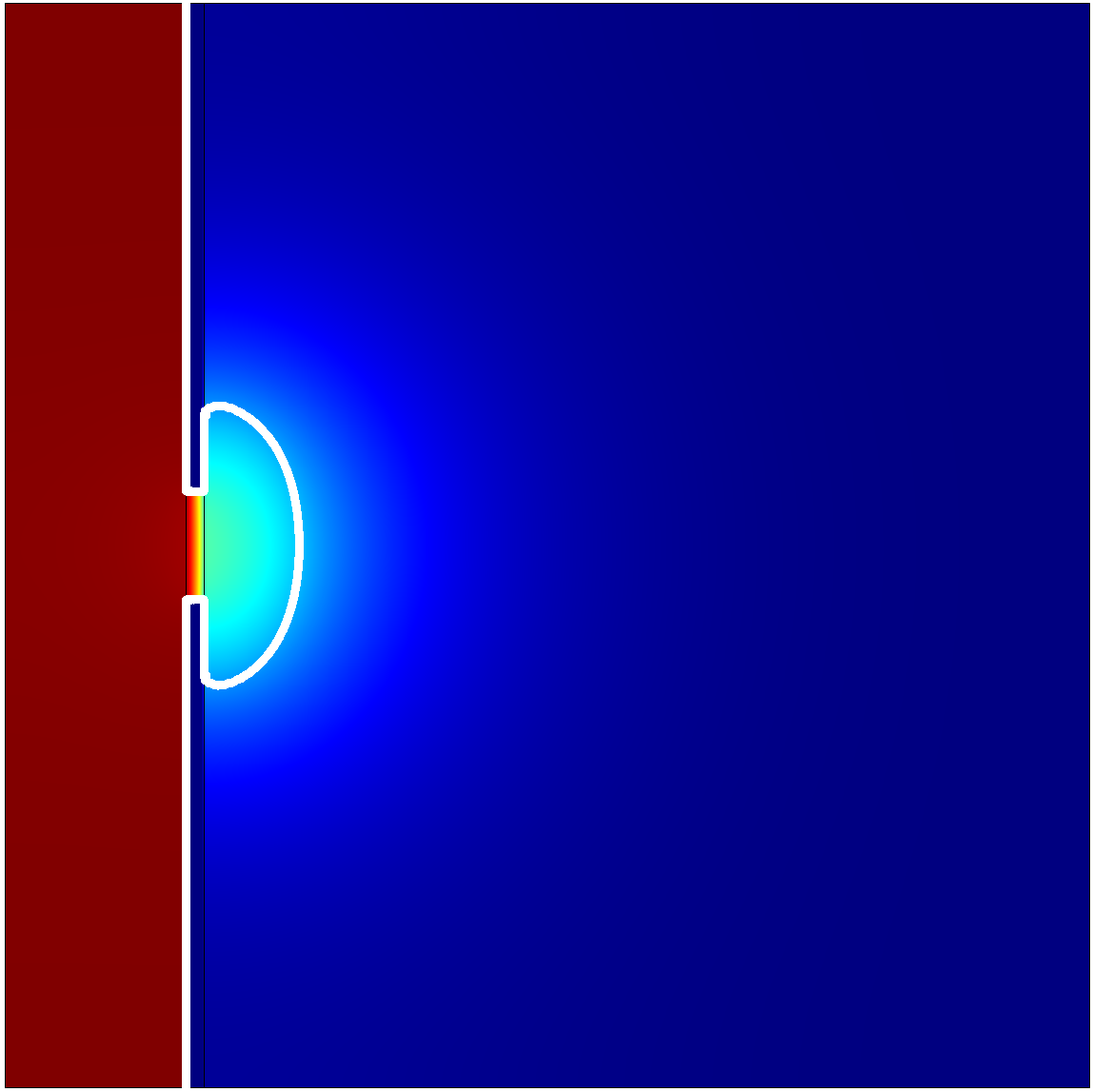}}
\subfigure[$t=20$]{\label{fig:2Dc_eps0p1}\includegraphics[width=0.22\textwidth]{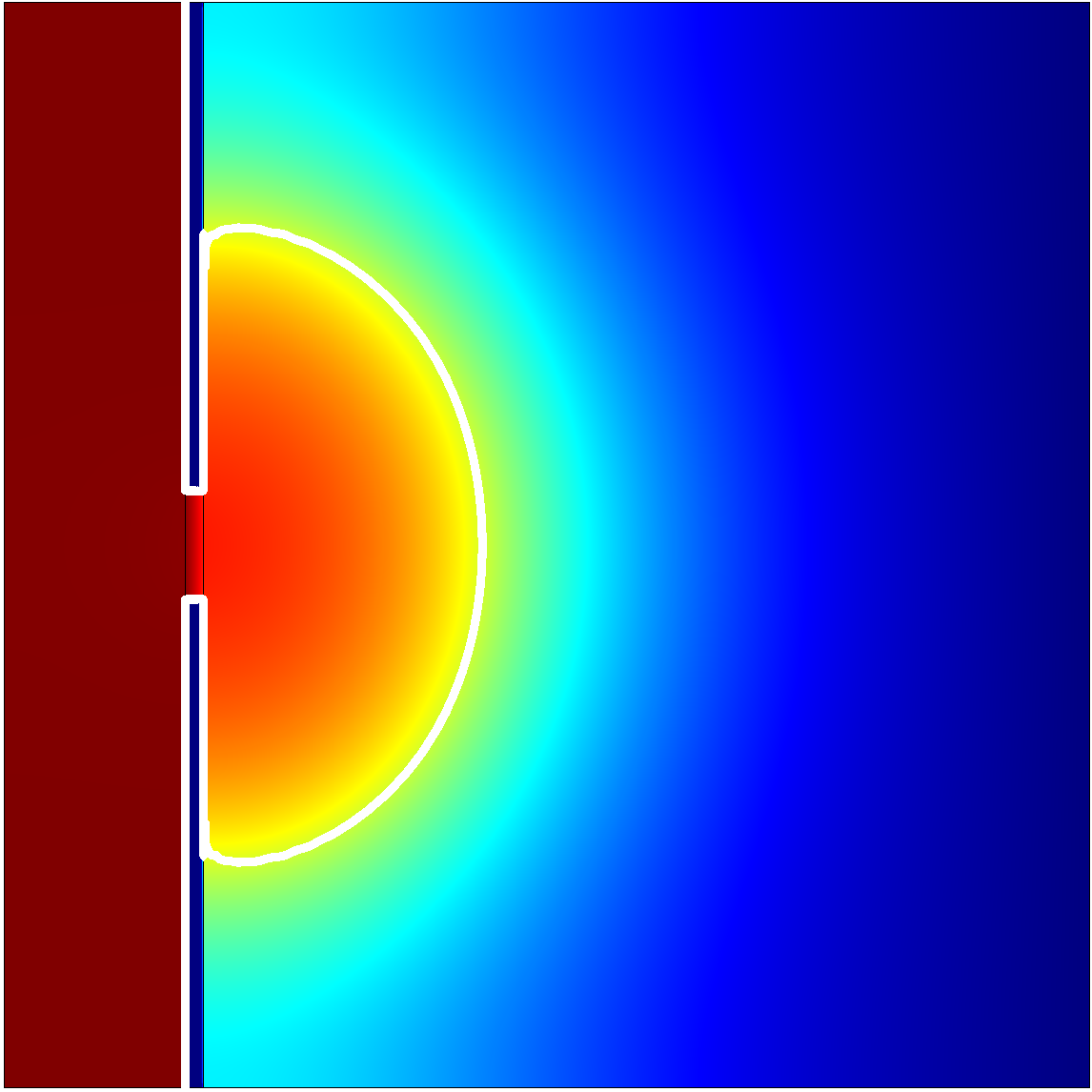}}
\subfigure[$t=30$]{\label{fig:2Dd_eps0p1}\includegraphics[width=0.22\textwidth]{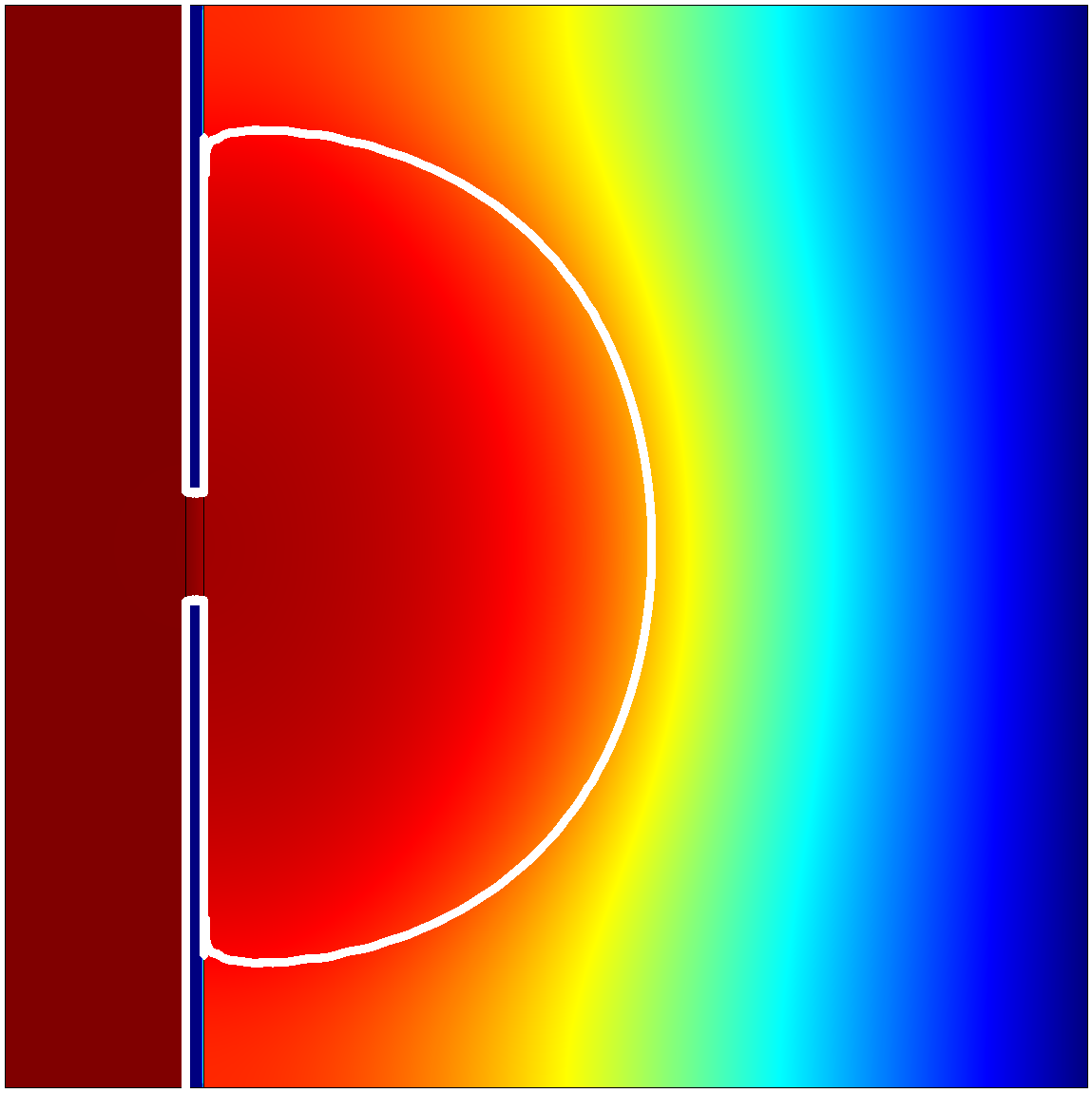}}
\centering     
\subfigure[$t=5$]{\label{fig:2Da}\includegraphics[width=0.22\textwidth]{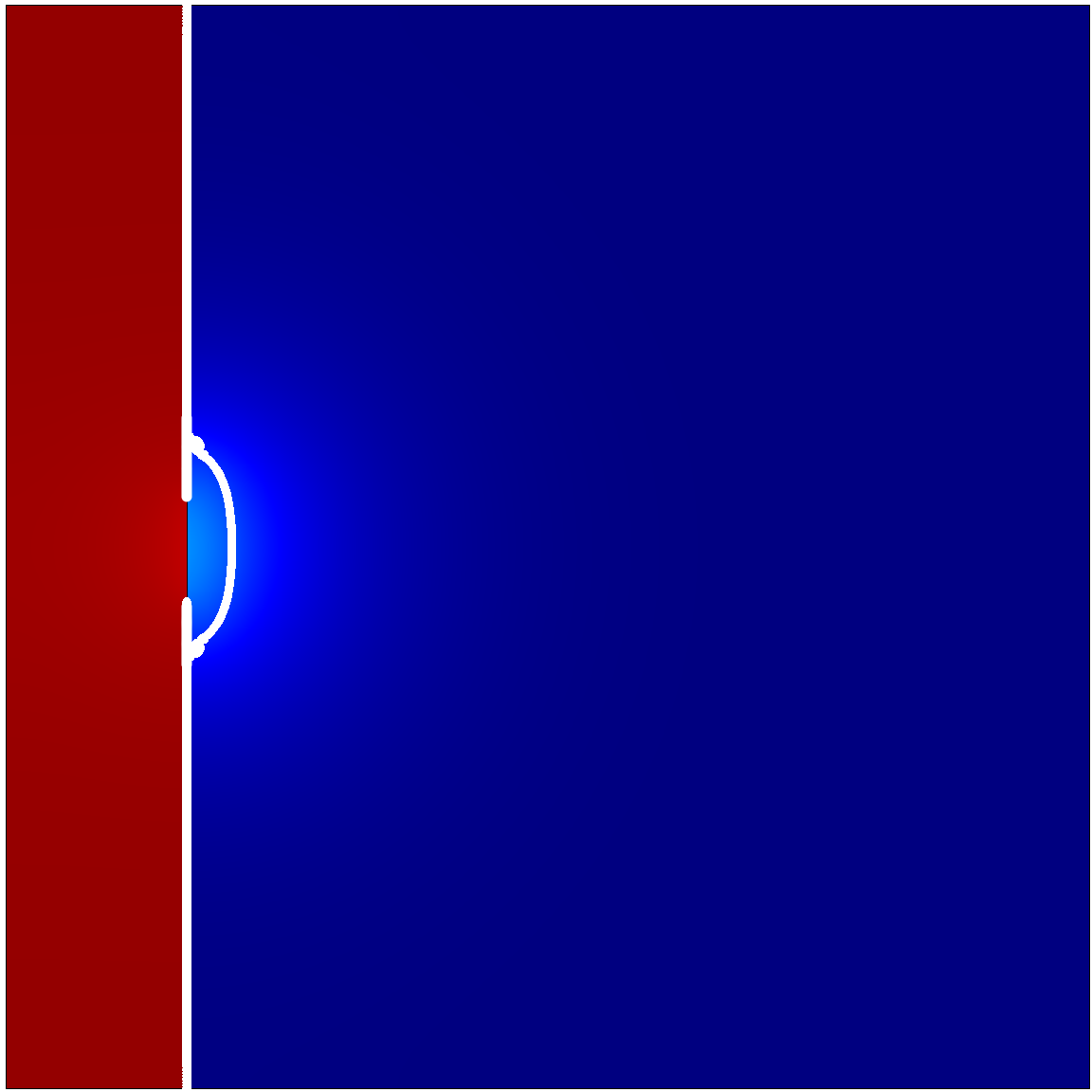}}
\subfigure[$t=10$]{\label{fig:2Db}\includegraphics[width=0.22\textwidth]{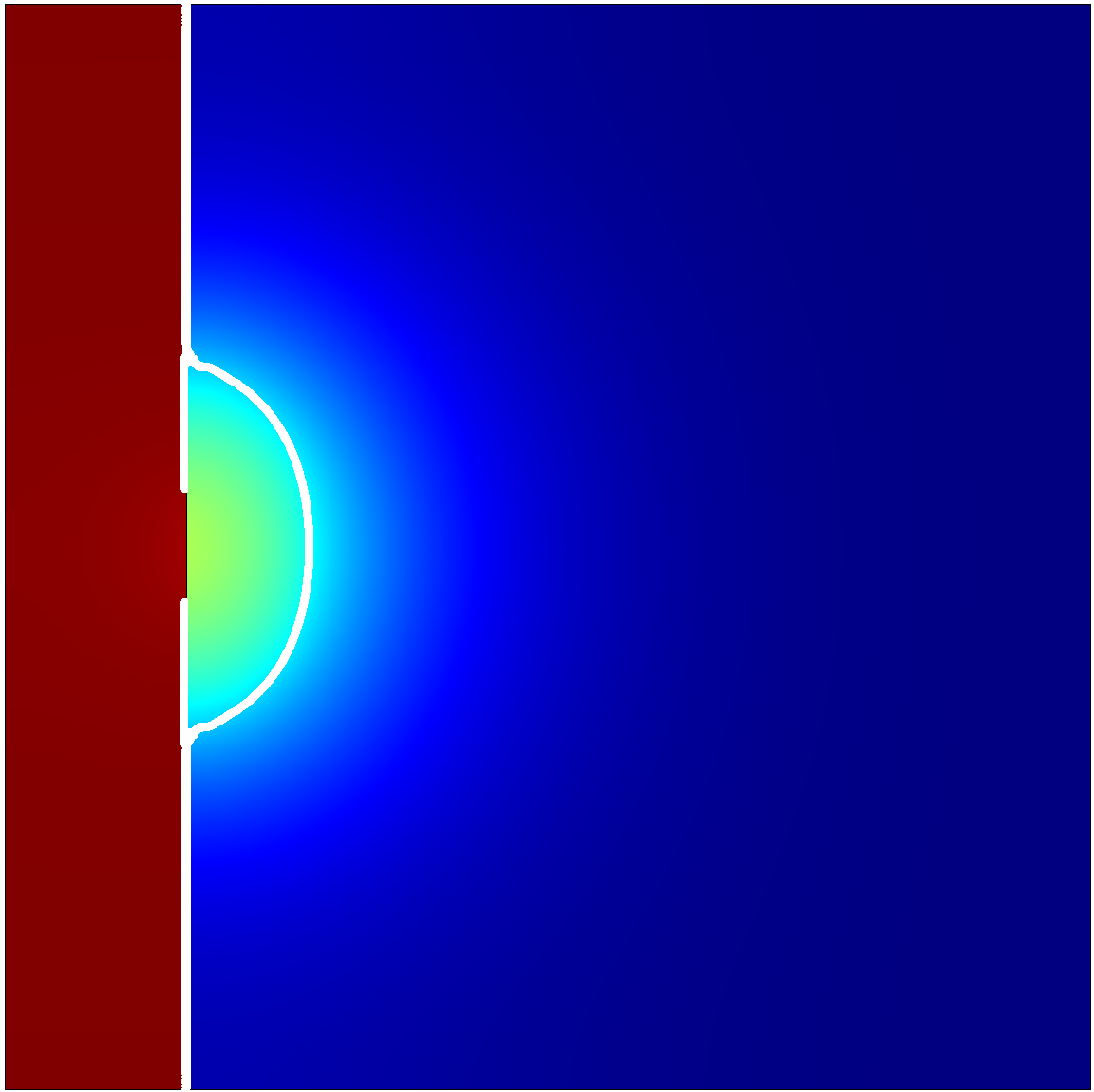}}
\subfigure[$t=20$]{\label{fig:2Dc}\includegraphics[width=0.22\textwidth]{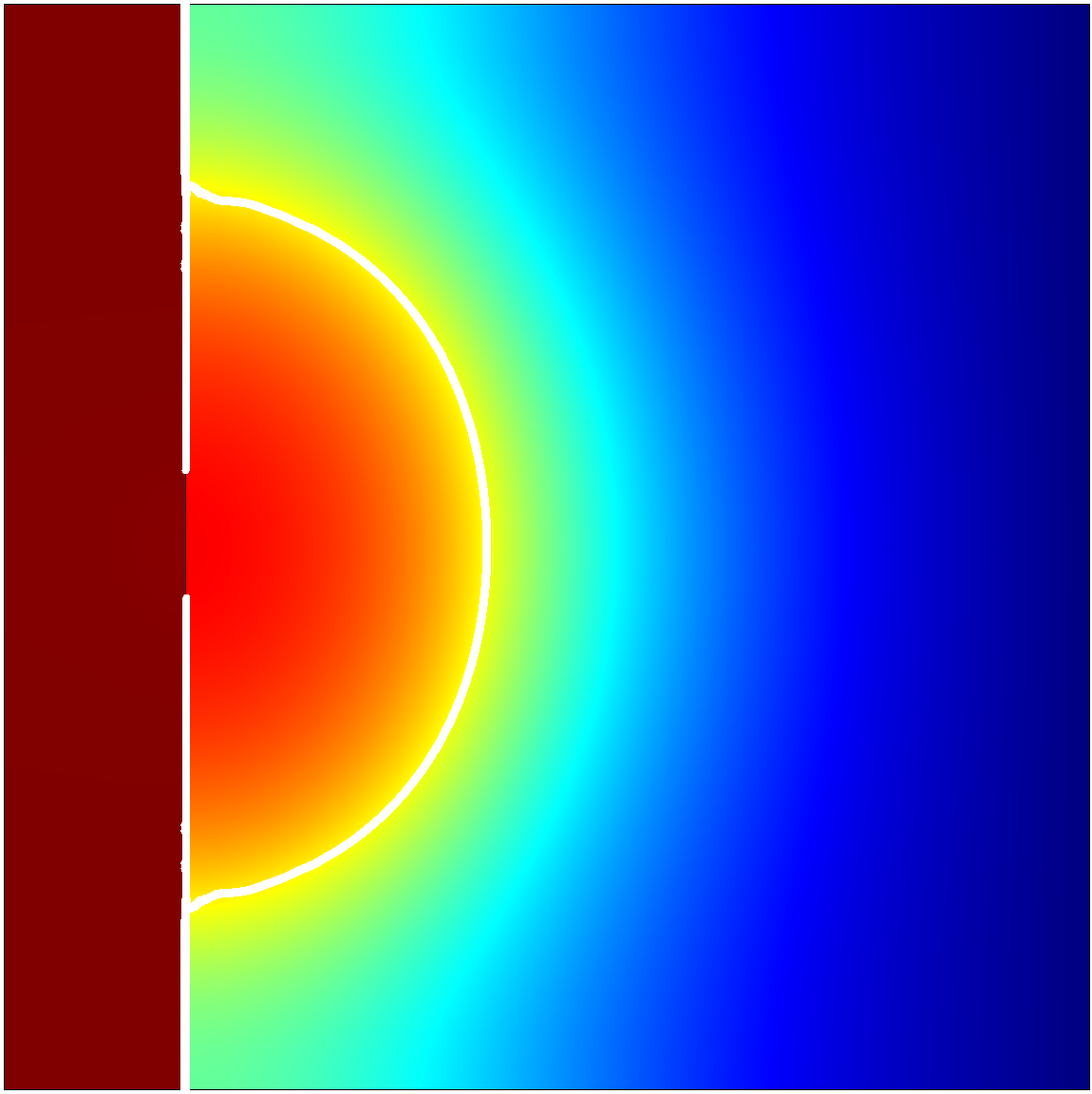}}
\subfigure[$t=30$]{\label{fig:2Dd}\includegraphics[width=0.22\textwidth]{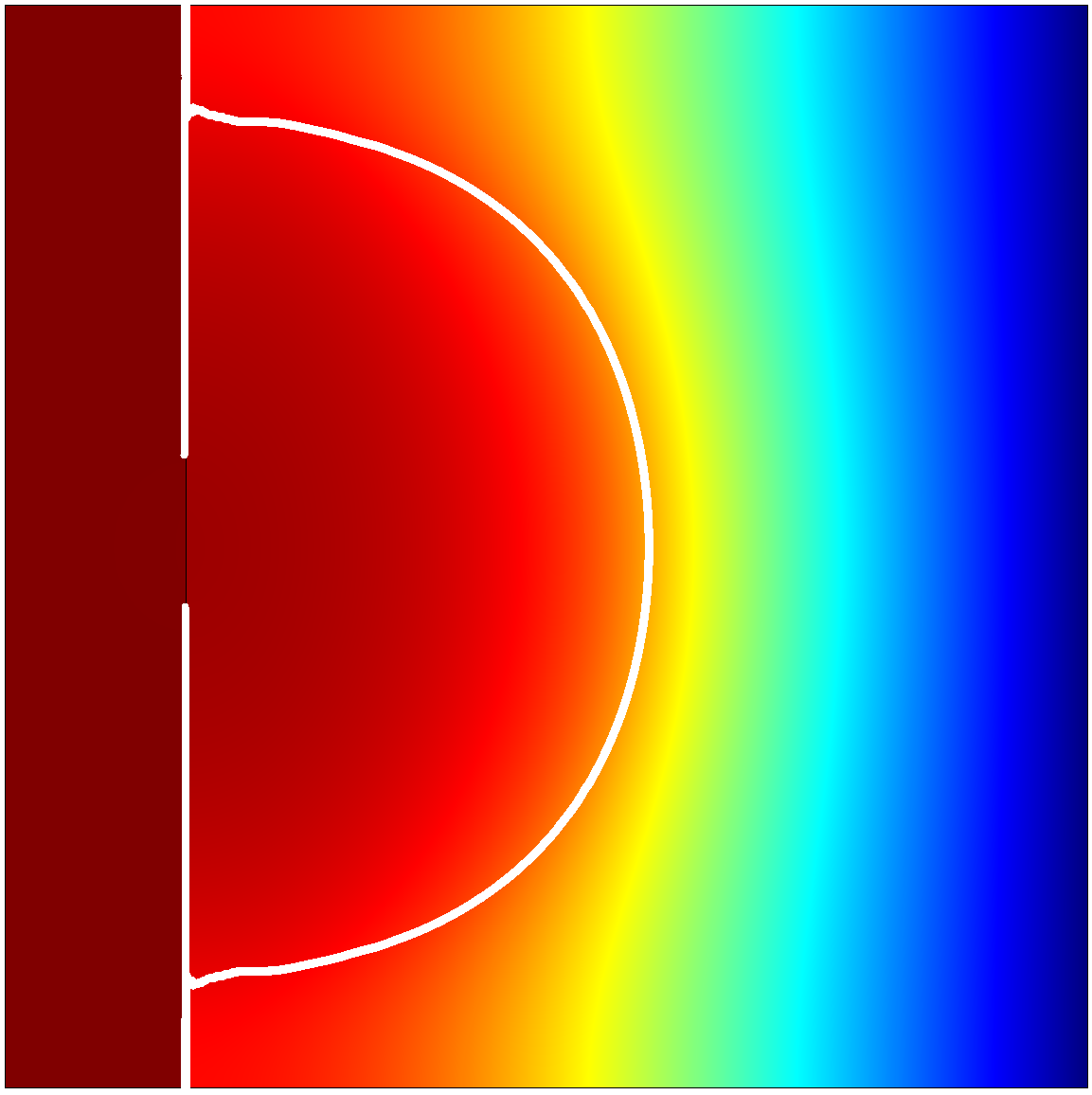}}
\caption{{\bf Numerical simulation of cancer cell invasion through the basement membrane.}
{\bf (a)}-{\bf (d)} Numerical solutions to the thin layer problem with $\e=0.1$. The different panels display the cell volume fraction $\rho_{i \e}(t,\x)$ with $i=1,2,3$ at successive non-dimensionalised time instants. 
{\bf (e)}-{\bf (h)}  Numerical solutions to the effective interface problem. The different panels display the cell volume fraction $\tilde{\rho}_{i}(t,\x)$ with $i=1,3$ at successive non-dimensionalised time instants. The colour scale ranges from blue (corresponding to $0.5$) to red (corresponding to $1$). The white curves are isolines that track the region of space occupied by cancer cells. To construct numerical solutions, we impose zero Neumann boundary condition on the left outer boundary and on the upper and lower boundaries, whereas a Dirichlet boundary condition is prescribed on the right outer boundary. The cells are uniformly distributed across the spatial domain at $t=0$, that is, we impose the following initial conditions $\rho_{i \e}(0,\x) := \rho_0$ for all $\x \in {\cal D}_{i \e}$ with $i=1,2,3$, and we consider a biological scenario whereby cancer cells are initially confined to the subdomain ${\cal D}_{1 \e}$ by making the assumption that $\varphi_\e(0, x,\cdot) := -x$. We choose the parameter values $\rho_0 = 0.5$, $\tilde \mu_{1} = \tilde \mu_{3} \equiv \bar{\mu}=0.5$ and $\bar{\mu}_2 = 0.1$.
\label{fig:2D_eps0p1}
}
\end{figure}

\begin{figure}[ht!]
\centering     
\subfigure[$t=0,5,10,15,20,25,30$]{\label{fig:conf2D_eps0p1}\includegraphics[width=0.325\textwidth]{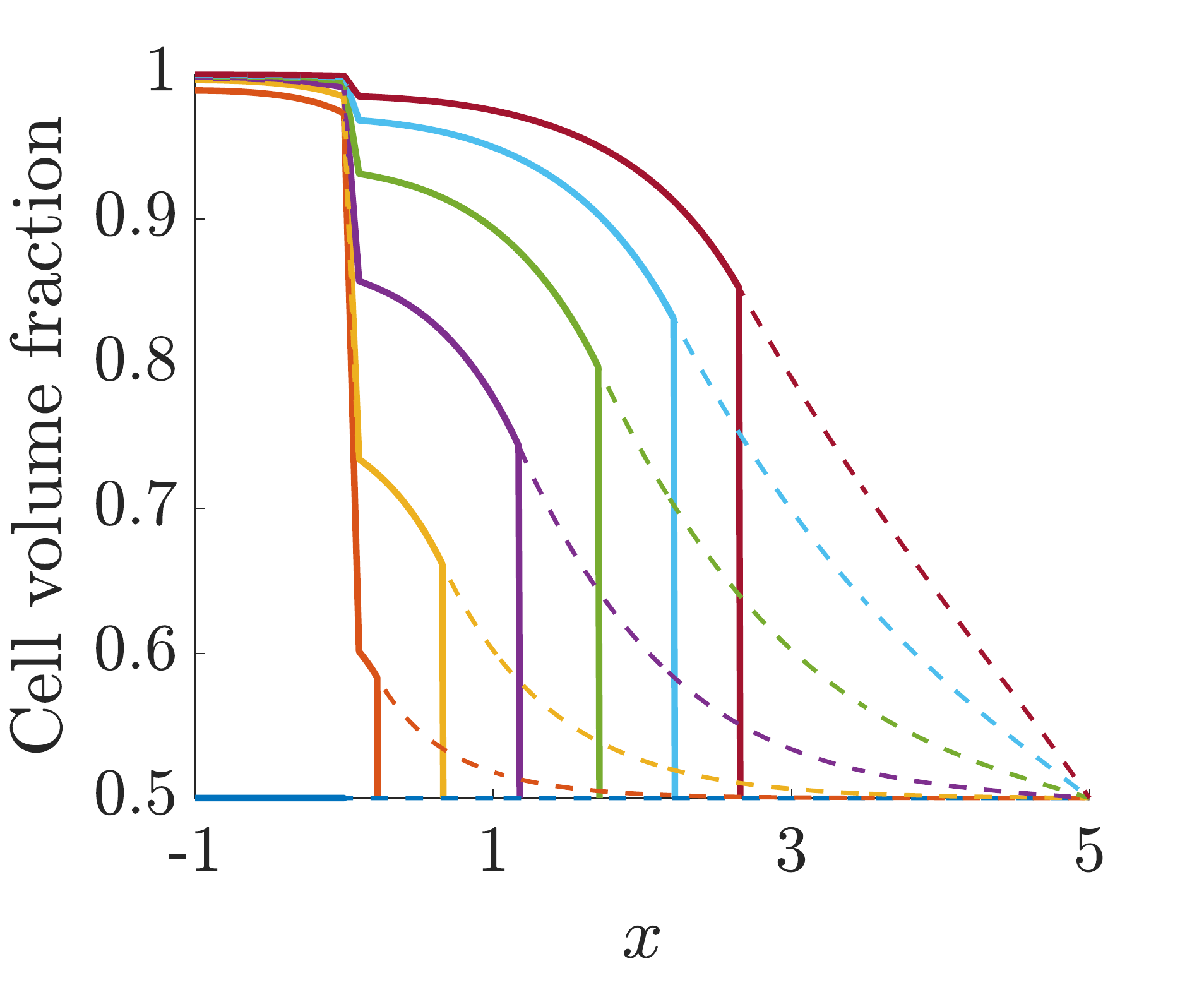}}
\subfigure[$t=0,5,10,15,20,25,30$]{\label{fig:conf2D_0}\includegraphics[width=0.325\textwidth]{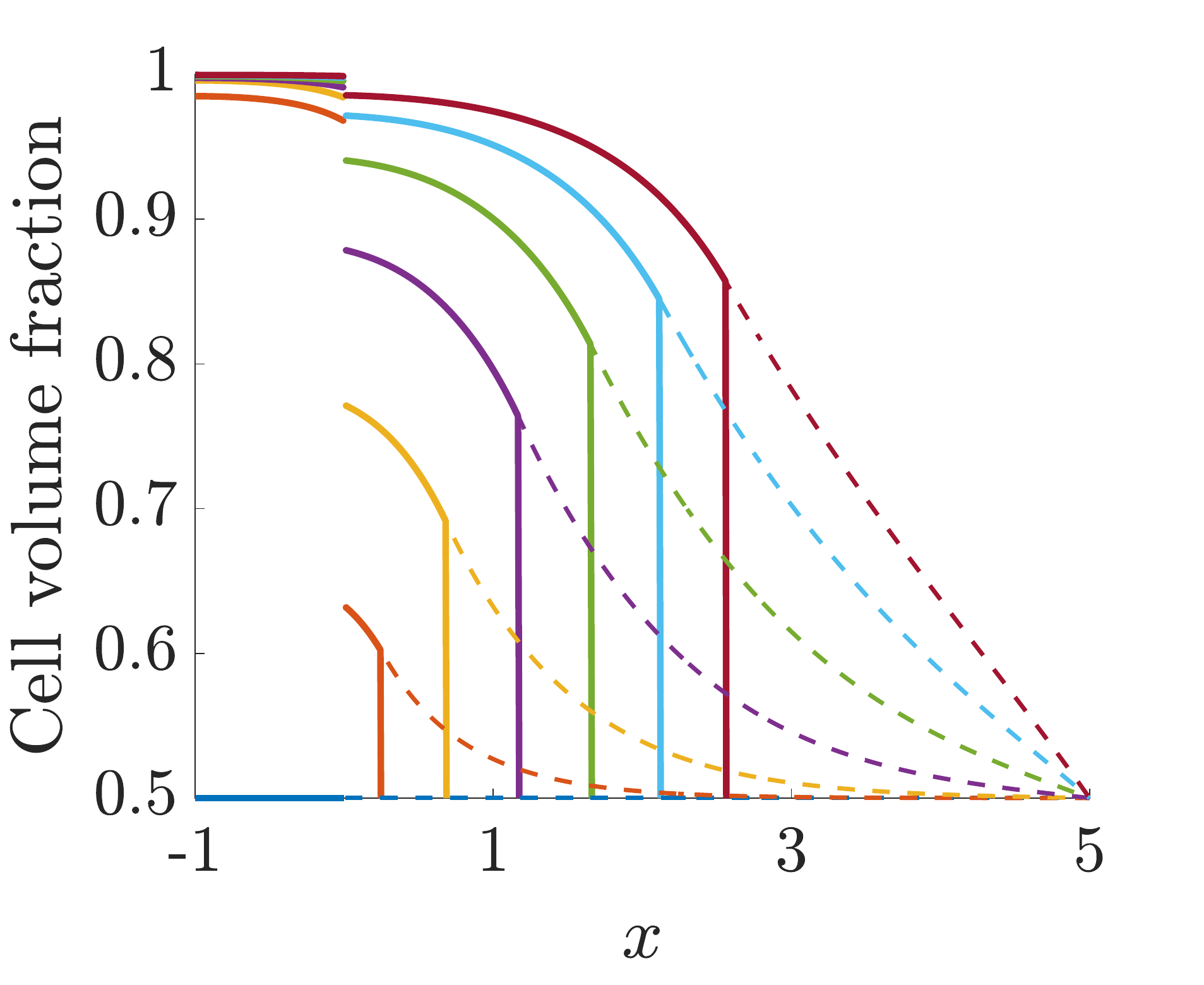}}
\subfigure[]{\label{fig:conf_diff}\includegraphics[width=0.325\textwidth]{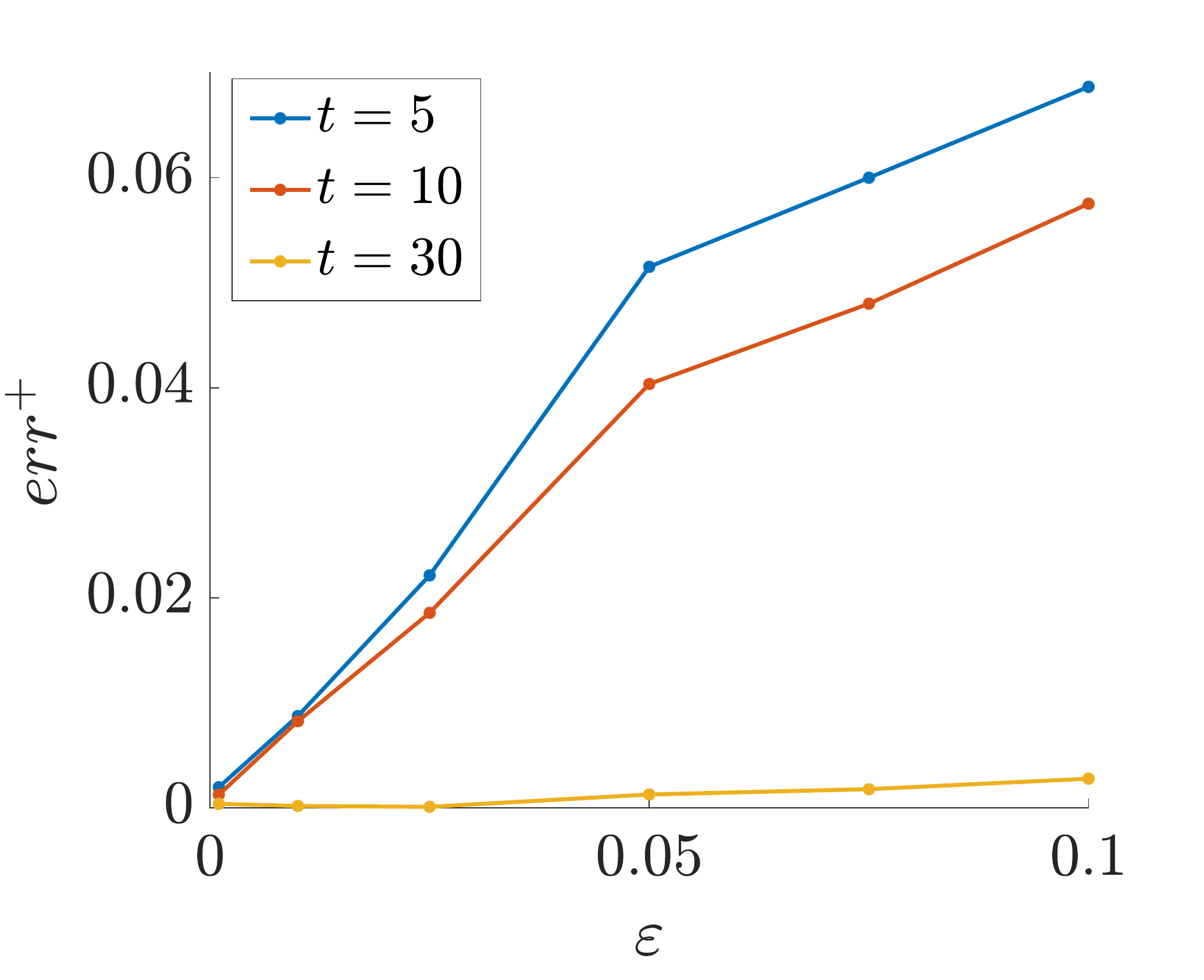}}
\caption{{\bf Numerical simulation of cancer cell invasion through the basement membrane.}
{\bf (a)} Spatio-temporal evolution of the volume fraction of cancer cells $\rho_{i \e}(t,x,0) \, H(\varphi_\e(t,x,0))$ (solid lines) and the volume fraction of healthy cells $\rho_{i \e}(t,x,0) \, \left(1-H(\varphi_\e(t,x,0))\right)$ (dashed lines) for the thin layer problem, with $i=1,2,3$. {\bf (b)} Spatio-temporal evolution of the volume fraction of cancer cells $\tilde \rho_{i}(t,x,0) \, H(\tilde \varphi(t,x,0))$ (solid lines) and the volume fraction of healthy cells $\tilde \rho_{i}(t,x,0) \, \left(1-H(\tilde \varphi(t,x,0))\right)$ (dashed lines) for the effective interface problem, with $i=1,3$. {\bf (c)}~Relative error between the numerical solutions to the thin layer problem at the point $(\e,0)$ and the numerical solutions to the effective interface problem at the point $(0^+, 0)$ (i.e. the quantity $err^+=|\rho_{3 \e}(t,\e,0) - \tilde \rho_3(t,0^+,0)|/\tilde \rho_3(t,0^+,0)$) as a function of $\e$, at successive time instants. The relative error at the point $(0^-,0)$ is not reported as it was smaller than $5 \times 10^{-3}$ for all $t$ and $\e$. 
\label{fig:confronto2D}}
\end{figure}

The numerical results obtained are summarised by the plots in Figs.~\ref{fig:2D_eps0p1} and~\ref{fig:confronto2D}. The plots on the top line of Fig.~\ref{fig:2D_eps0p1} display the numerical solutions to the thin layer problem with $\e=0.1$ at different time instants. The numerical solutions to the effective interface problem at the same time instants are displayed in the plots on the bottom line. The discrepancy between the solutions to the thin layer problem and the solutions to the effective interface problem decays over time as the invasion front of cancer cells moves away from the basement membrane, which is represented either by the subdomain ${\cal D}_{2 \e}$ or by the effective interface $\tilde \Sigma_{13}$. \red{This is further clarified by the plots in Fig.~\ref{fig:confronto2D}. In particular, the curves reported in Fig.~\ref{fig:conf_diff} indicate that the relative error between the numerical solution to the thin layer problem at the point $(\e, 0)$ and the numerical solution to the effective interface problem at the point $(0^+,0)$ decays over time. Moreover, in agreement with the formal results established by Proposition~\ref{Th1}, the relative error decays as $\e \to 0$. }

\subsection{Numerical simulation of ovarian cancer invasion}
\label{secovary}
In this section, we apply the formal results established by Proposition~\ref{Th1} to the mathematical modelling of cell invasion dynamics in ovarian carcinoma. In particular, we simulate the metastatic journey of a cancer multicellular mass, from the initial growth inside the ovary to the invasion of the healthy tissue adjacent to the peritoneum, using an effective interface problem. 

For the sake of brevity, throughout this section we drop the tildes from all quantities and we work with dimensionless quantities, as specified in the previous subsection. In particular, we use the notation $\x = (x/L, y/L)$ to denote the spatial position non-dimensionalised with respect to the characteristic size $L>0$ of the region represented as the subdomain ${\cal D}_{1}$. 

\subsubsection{Biological background}
\label{bioback}
Ovarian carcinoma originates either inside the ovary or in the fallopian tube. This type of cancer is known to invade the surrounding tissues and to metastasise both by direct extension and by cell detachment from the primary tumour \cite{lengyel2010ovarian}. The latter process of metastasis formation is peculiar to ovarian carcinoma and allows cancer cells to spread into the peritoneal cavity, to invade adjacent peritoneal tissues and, ultimately, to reach distant organs. Such a process encompasses multiple layers of complexity, which represents one of the main reasons why the metastatic behaviour of ovarian cancer cells remains poorly understood.

The detachment of ovarian cancer cells from the primary tumour starts with the destruction of the basement membrane underling the ovarian capsule (\emph{i.e.} the ovarian surface epithelium) \cite{ahmed2007epithelial}. Cancer cells can subsequently break through the ovarian capsule as single cells or, more frequently, as spheroid-like aggregates. Such multicellular masses grow and passively move until they reach the walls of the peritoneal cavity -- which represent the common site of disaggregation, dissemination and metastatic outgrowth for ovarian carcinoma \cite{giverso2010individual}. 

The cancer cells that reach the walls of the cavity can attach to the mesothelial cells that constitute the peritoneal lining and, by secreting MMPs \cite{lengyel2010ovarian}, they can degrade the basement membrane underling the mesothelium and cleave cell-cell adhesion molecules (\emph{e.g.} N-cadherins) that hold mesothelial cells together \cite{lengyel2010ovarian}. This leads to the retraction of mesothelial cells at the cancer cells' attachment sites and promotes the formation of foci of invasion, which enable the ovarian cancer cells to invade the healthy tissue adjacent to the peritoneum and form secondary tumours \cite{giverso2010individual}. 

\subsubsection{Mathematical model}
In adult human females, the ovarian capsule consists of a single layer of epithelial cells and the peritoneal lining is constituted by a monolayer of mesothelial cells \cite{ahmed2007epithelial}. Hence, the thickness of the ovarian capsule and the peritoneal lining is small compared to the characteristic size of the ovary and of the peritoneal cavity. For this reason, we represent both the ovarian capsule and the peritoneal lining, along with the underling basement membranes, as two thin porous membranes. Moreover, using the formal results established by Proposition~\ref{Th1}, we model each thin porous membrane as an effective interface.

\begin{figure}
\centering     
\includegraphics[width=0.35\textwidth]{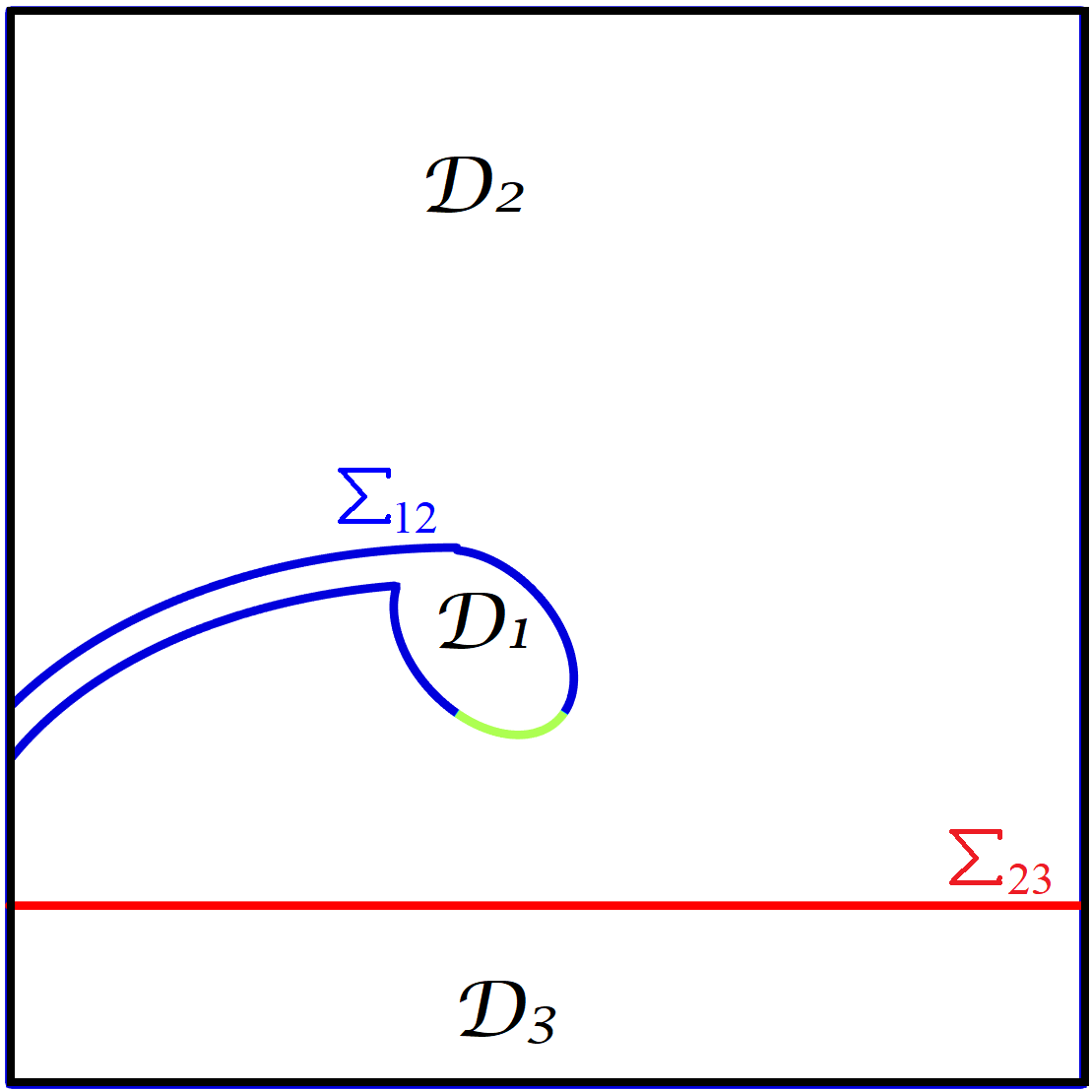}
\caption{{\bf Spatial domain used in the numerical simulation of ovarian cancer invasion.}
The subdomain ${\cal D}_1$ corresponds to the ovary, the subdomain ${\cal D}_2$ represents the peritoneal cavity and the subdomain ${\cal D}_3$ models the healthy tissue adjacent to the peritoneum. The effective interfaces $\Sigma_{12}$ and $\Sigma_{23}$ represent, respectively, the ovarian capsule and the peritoneal lining. The part of the ovarian capsule highlighted in green (i.e. $\Sigma_p \subset \Sigma_{12}$) is assumed to be permeable to cancer cells.\label{fig:geometria_ovaio}}
\end{figure}

On the basis of these observations, considering a two-dimensional scenario, we represent the ovary, the peritoneal cavity and the healthy tissue adjacent to the peritoneum as three distinct spatial subdomains ${\cal D}_1$, ${\cal D}_2$ and ${\cal D}_3$ separated by the effective interfaces $\Sigma_{12}$ (\emph{i.e.} the ovarian capsule along with the underlying basement membrane) and $\Sigma_{23}$ (\emph{i.e.} the peritoneal lining along with the underlying basement membrane) -- \emph{cf.} respectively, the blue curve and the red line in Fig.~\ref{fig:geometria_ovaio}. We focus on the biological scenario whereby there is a part of the ovarian capsule that is damaged and thus permeable to cancer cells. We identify such a region with a subset $\Sigma_p$ of the effective interface $\Sigma_{12}$ (\emph{cf.} the green line in Fig.~\ref{fig:geometria_ovaio}).

Letting the function $\rho_i(t,\x)$ model the cell volume fraction at position $\x \in {\cal D}_i$ and time $t \geq 0$, we describe the spatio-temporal evolution of the cells through the effective interface problem~\eqref{P0}, with $i = \left\lbrace 1,2,3\right\rbrace$, posed on the spatial domain illustrated in Fig.~\ref{fig:geometria_ovaio}. Similarly to Section~\ref{num2d}, we define $f(\rho_i)$ according to \eqref{1ddb} and we let ovarian cancer cells proliferate in all subdomains ${\cal D}_i$, with $i=\left\lbrace 1,2,3 \right\rbrace$, according to the net growth rate $\Gamma_i \equiv \Gamma(\varphi,\rho_i)$ given by the logistic law~\eqref{Gamma}. The evolution of the function $\varphi(t, \x)$ is governed by Eq.~\eqref{eq:levelset0} posed on the spatial domain illustrated in Fig.~\ref{fig:geometria_ovaio} and subject to the continuity condition~\eqref{eq:levelset0cont} on $\Sigma_{12}$ and $\Sigma_{23}$.

We make the prima facie assumption that the effective mobility coefficient ${\mu}_{12}$ is a given function of ${\bf x}$ and does not depend on $t$. On the other hand, on the basis of the biological facts discussed in Section~\ref{bioback}, we let the effective mobility coefficient ${\mu}_{23}$ be a function of the local concentration of MMPs $c(t,\x)$, which can vary across space and time. In particular, using a modelling strategy similar to that proposed by Gallinato \emph{et al.}~\cite{Gallinato} and Giverso \emph{et al.}~\cite{giverso2018nucleus}, we define ${\mu}_{23}$ as
\begin{equation} \label{eq:mu23tilderis}
{\mu}_{23}(t,\x) \equiv {\mu}_{23}(c(t,\x)) := \bar{\mu}_{23} \dfrac{ \left(c(t,\x) - 1 \right)_+}{K_c + (c(t,\x)-1)}, \quad \bar{\mu}_{23}>0, \; K_c >0.
\end{equation}
\red{A detailed derivation of definition~\eqref{eq:mu23tilderis} is provided in Section~S.3 of the Supplementary Material}. Denoting the restriction of the function $c$ to the subdomain ${\cal D}_i$ by $c_i$, we describe the dynamics of the concentration of MMPs through the following transmission problem 
\begin{equation}\label{eq:MMP}
\left\{
\begin{array}{ll}
\displaystyle{\frac{\partial c_i}{\partial t} = \gamma_c \, \rho_i \, H(\varphi)+ D_c \, \Delta  c_i} 
&{\rm in}\ {\cal D}_i, \quad\;\; i=1,2,3,\\[12pt]
D_c \, \nabla c_i \cdot {\bf n}_{ij} = D_c \, \nabla c_j \cdot {\bf n}_{ij}  &{\rm on}\ \Sigma_{ij}, \quad i=1,2, \quad j=i+1,\\[12pt]
[\![c]\!]  = 0&{\rm on}\ \Sigma_{ij}, \quad i=1,2, \quad j=i+1,
\end{array}
\right.
\end{equation}
where the parameter $\gamma_c >0$ is the rate at which cancer cells release MMPs and the parameter $D_c >0$ is the diffusivity of MMPs. Notice that the transmission conditions in~\eqref{eq:MMP} are such that the MMP concentration $c(t,\x)$ and its flux are continuous across the effective interfaces $\Sigma_{12}$ and $\Sigma_{23}$. This is because the size of the MMP molecules is much smaller than the size of the pores of the membranes (\emph{i.e.} the membranes are permeable to the MMP molecules). Alternatively, one could impose the classical Kedem-Katchalsky interface conditions on $\Sigma_{12}$ and $\Sigma_{23}$.

\subsubsection{Numerical solutions}
\begin{figure}[h!]
\centering     
\subfigure[$t=4$]{\label{fig:ovarya}\includegraphics[width=0.22\textwidth]{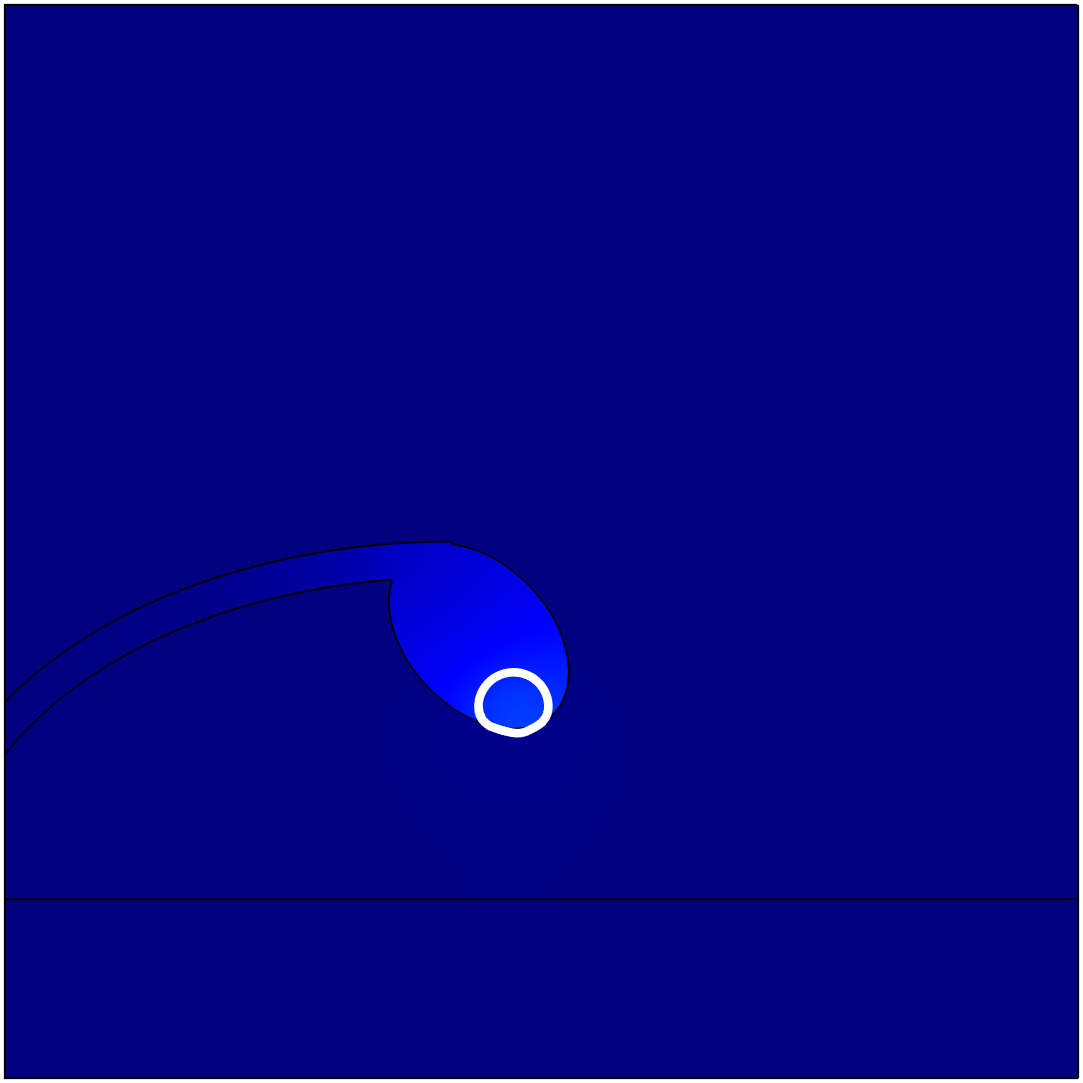}}
\subfigure[$t=8$]{\label{fig:ovaryb}\includegraphics[width=0.22\textwidth]{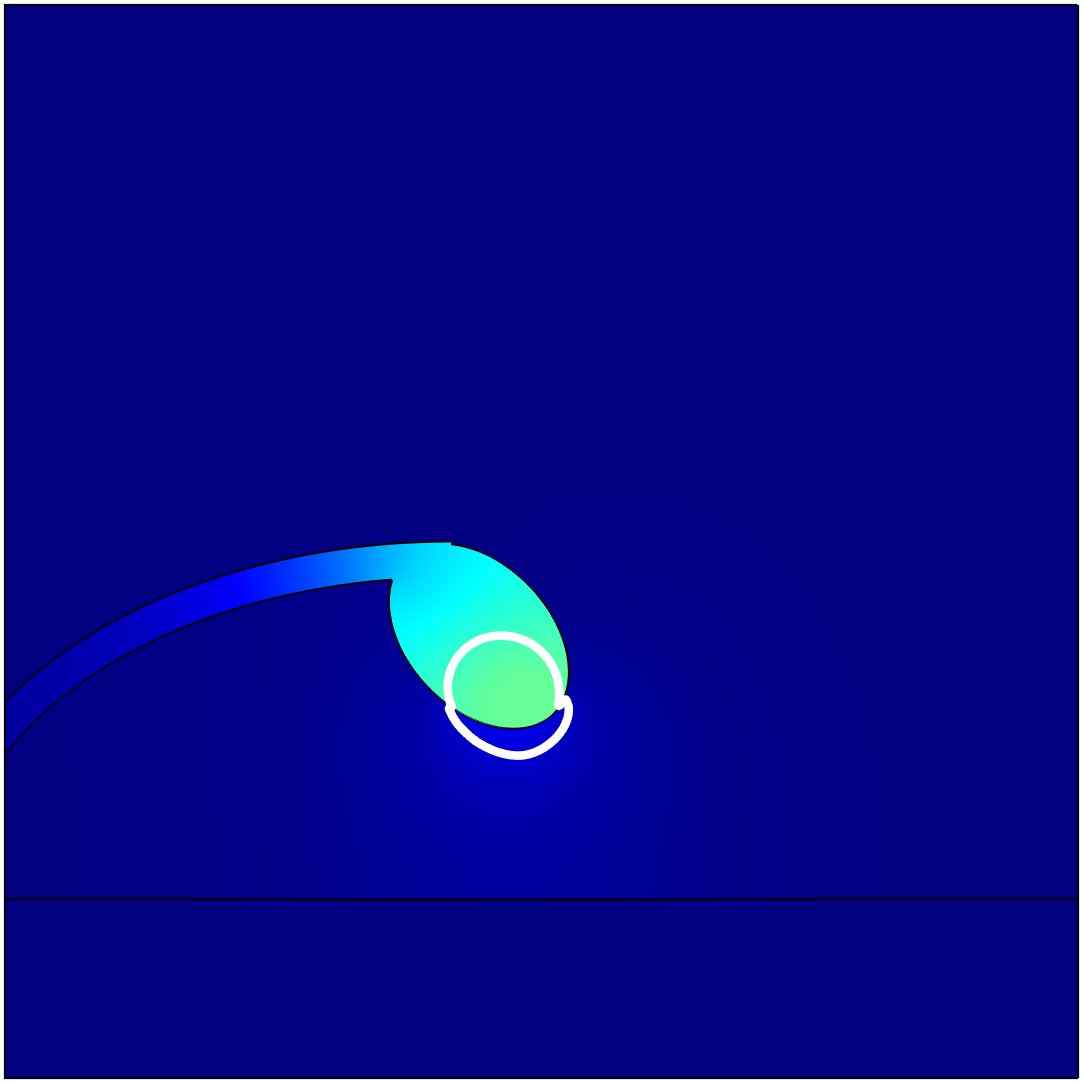}}
\subfigure[$t=14$]{\label{fig:ovaryc}\includegraphics[width=0.22\textwidth]{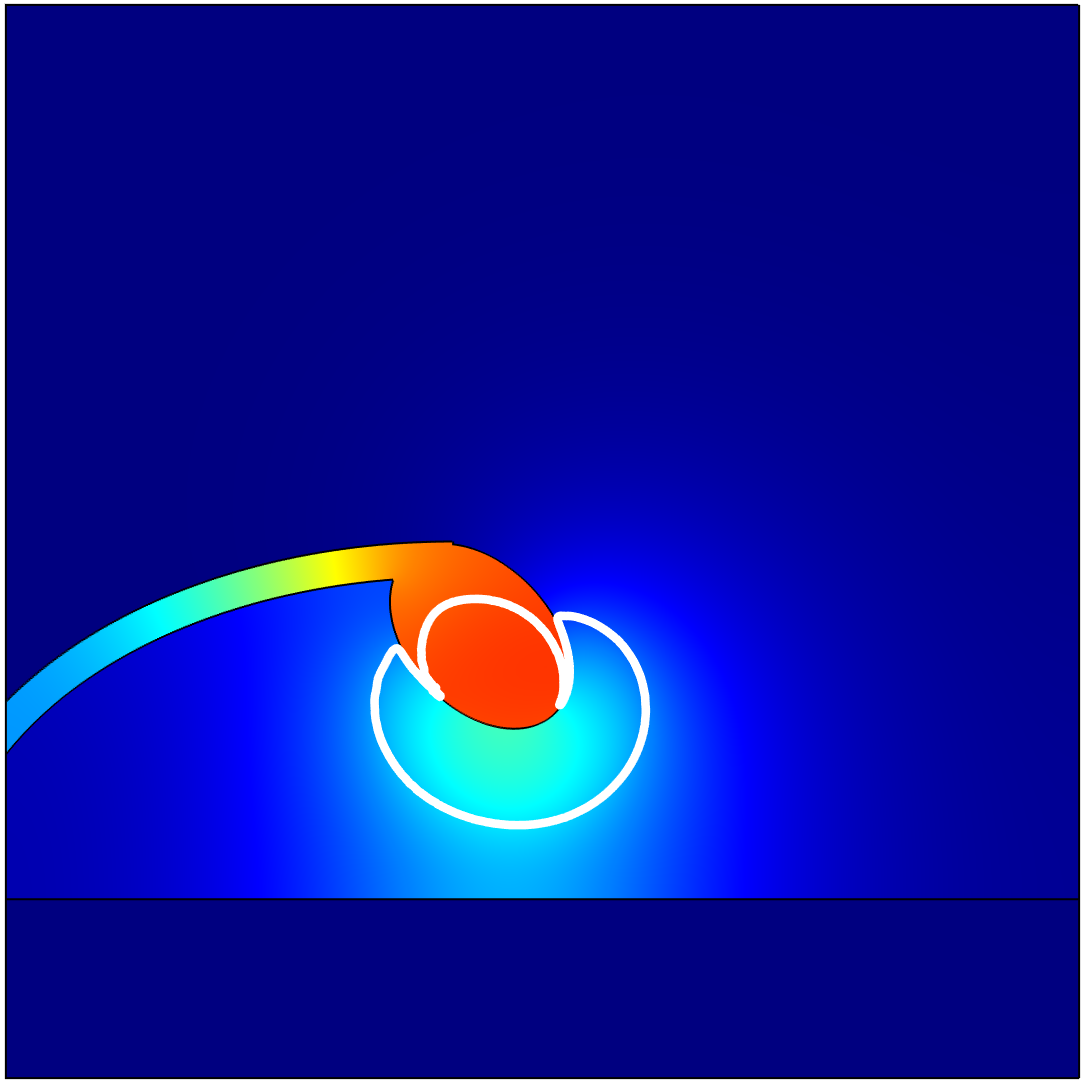}}
\subfigure[$t=20$]{\label{fig:ovaryd}\includegraphics[width=0.22\textwidth]{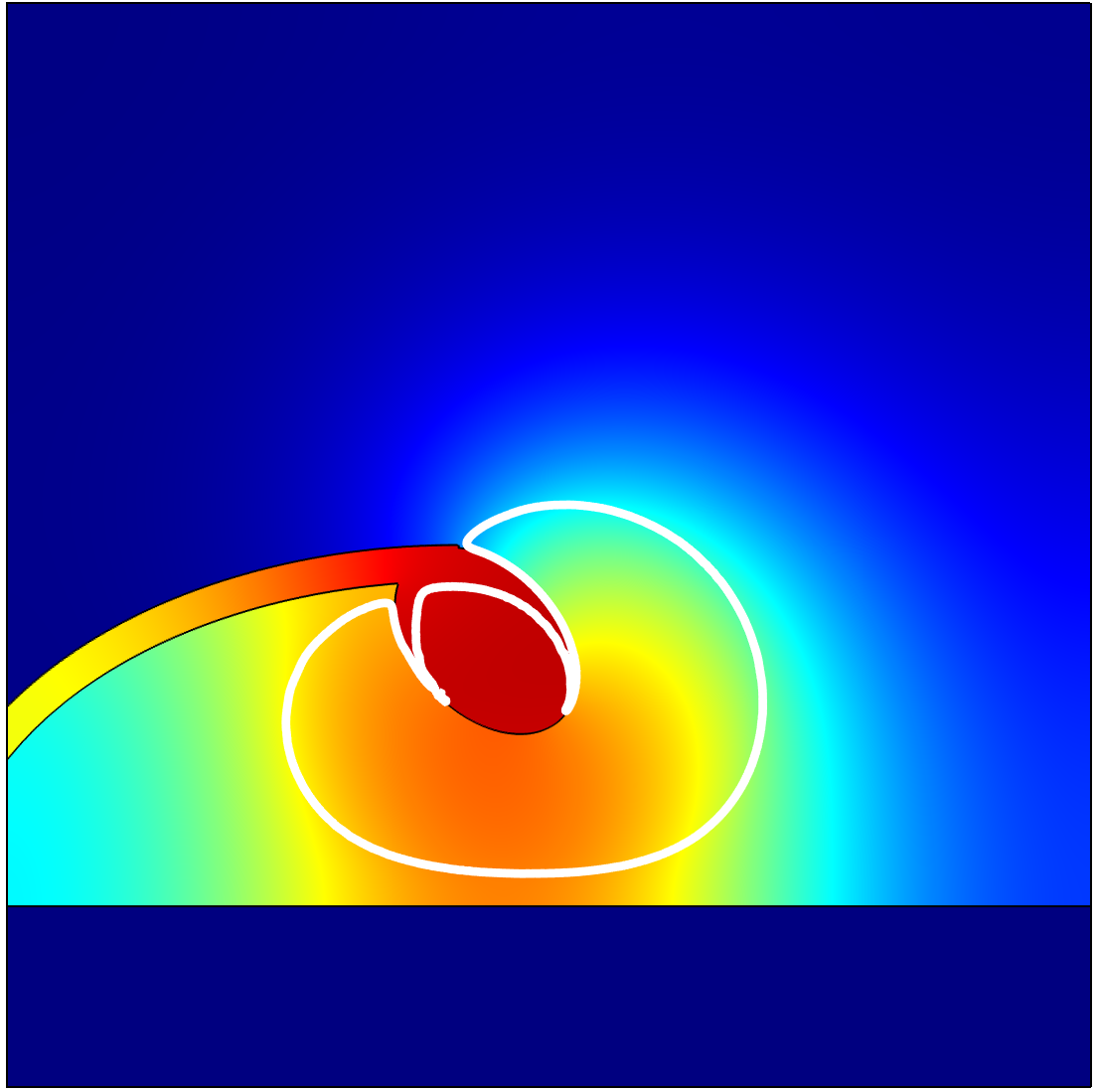}}
\subfigure[$t=28$]{\label{fig:ovarye}\includegraphics[width=0.22\textwidth]{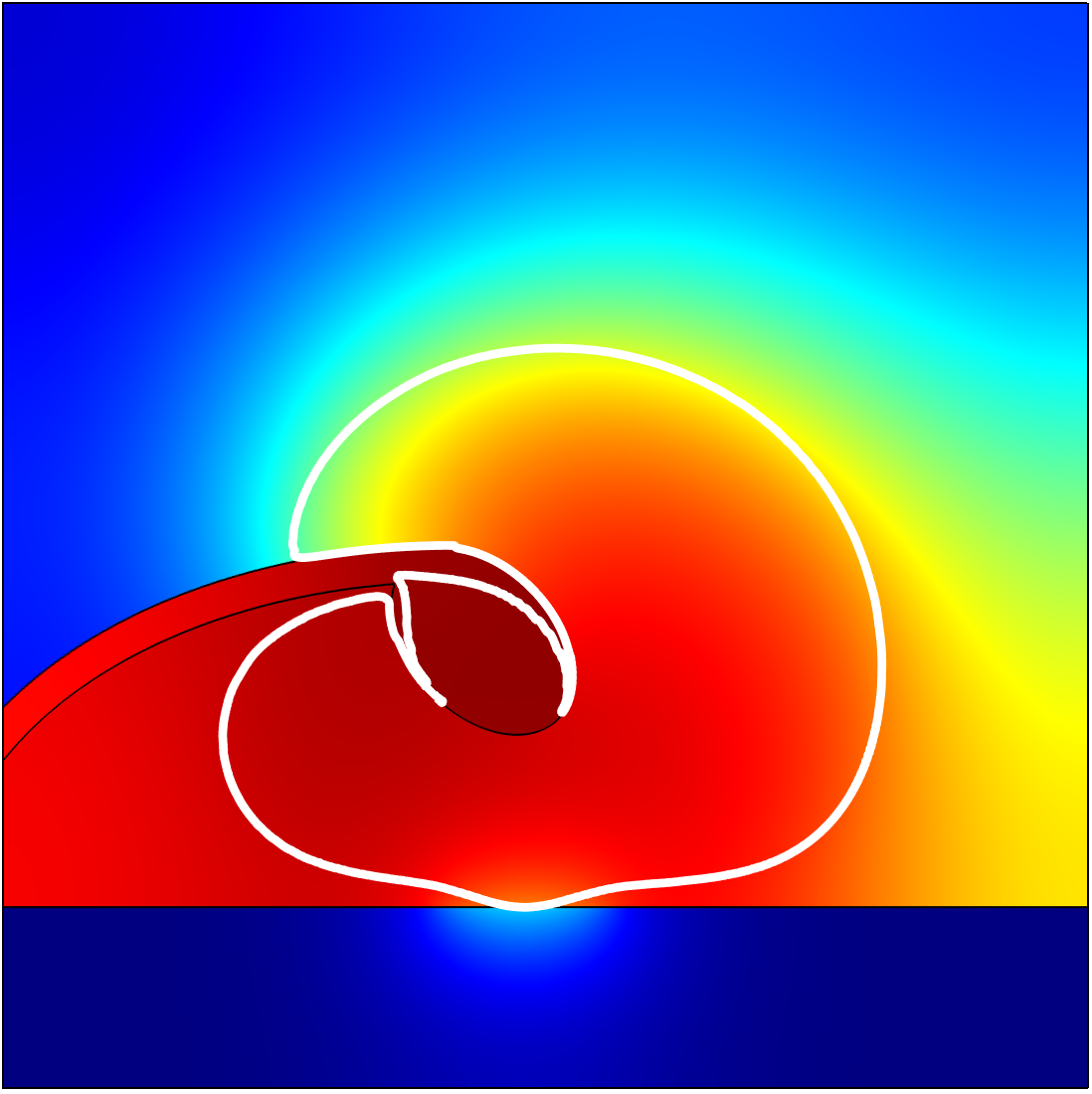}}
\subfigure[$t=30$]{\label{fig:ovaryf}\includegraphics[width=0.22\textwidth]{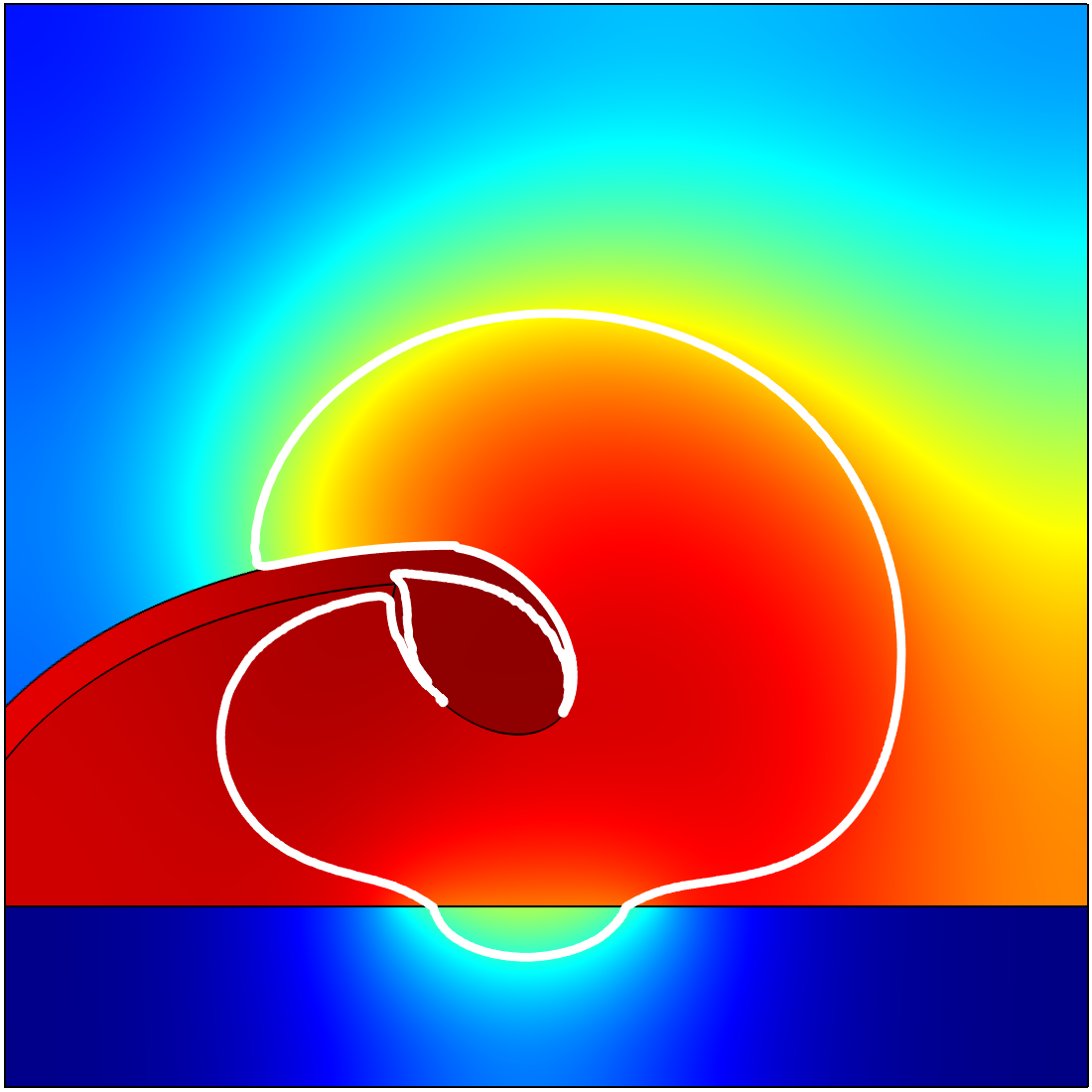}}
\subfigure[$t=32$]{\label{fig:ovaryg}\includegraphics[width=0.22\textwidth]{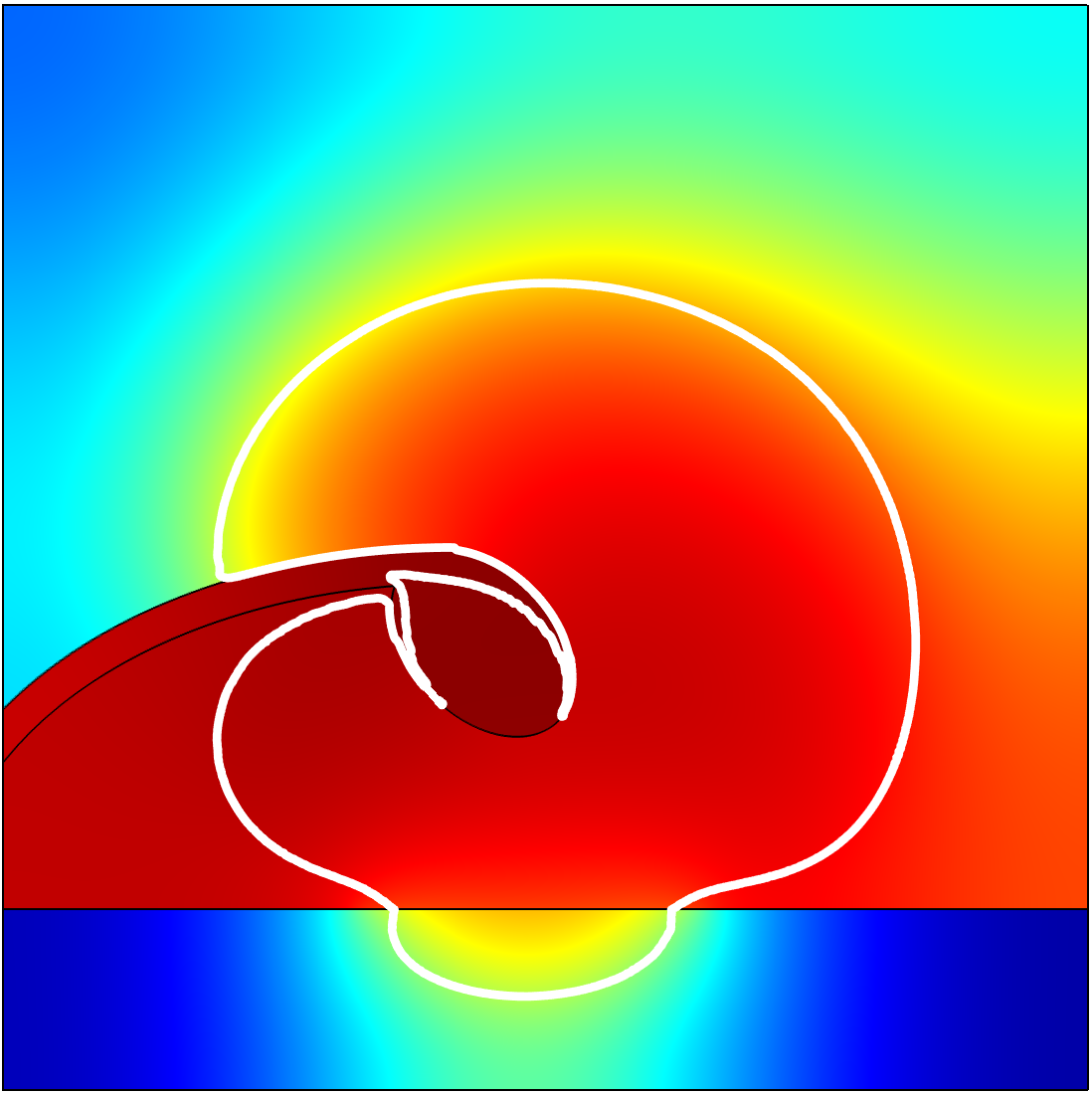}}
\subfigure[$t=36$]{\label{fig:ovaryh}\includegraphics[width=0.22\textwidth]{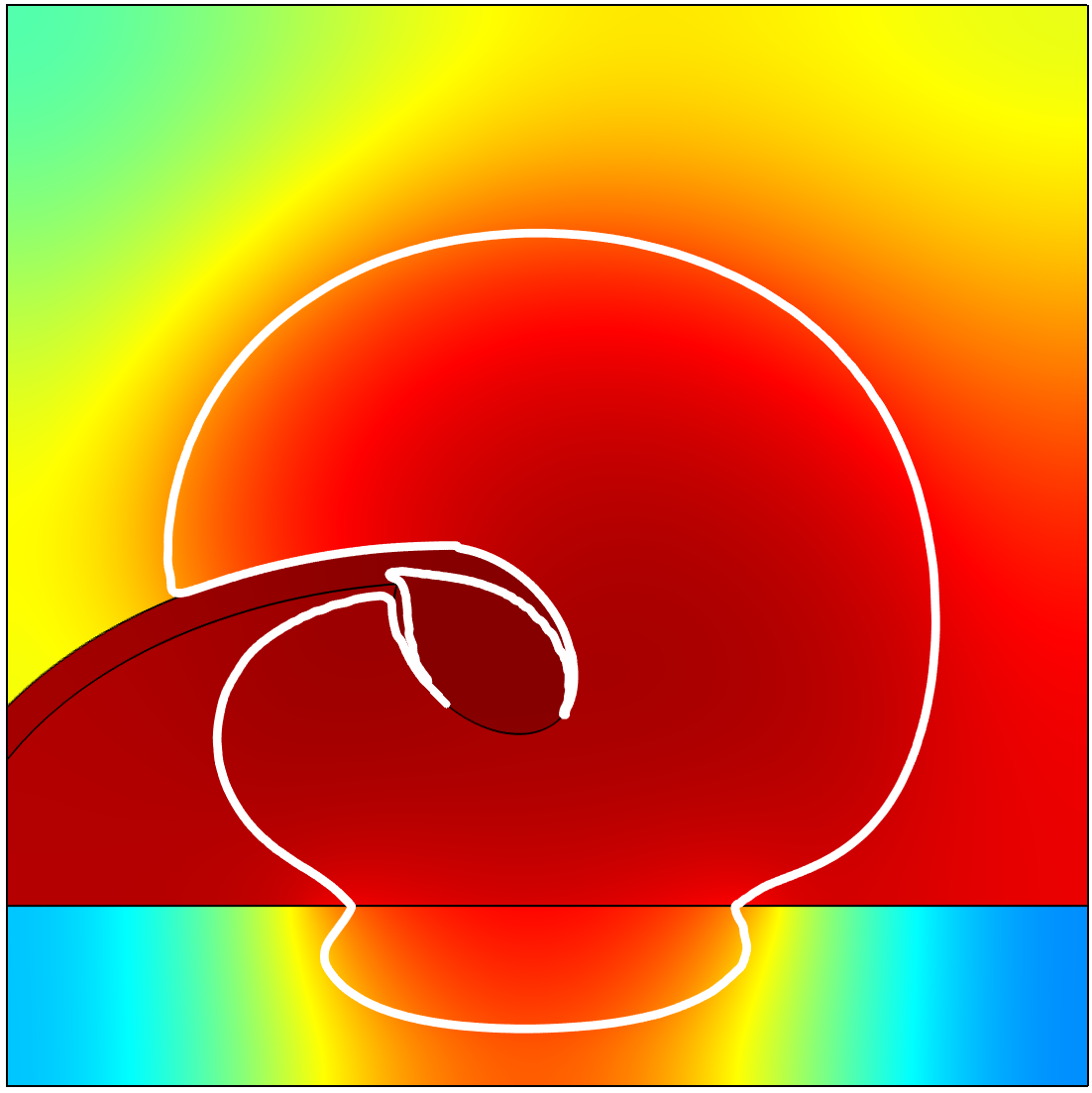}}
\caption{{\bf Numerical simulation of ovarian cancer invasion.} Numerical solutions to the transmission problem defined by the effective interface problem~\eqref{P0} posed on the spatial domain of Fig.~\ref{fig:geometria_ovaio} and coupled with the transmission problem~\eqref{eq:MMP}. The different panels display the cell volume fraction $\rho_i(t,\x)$ with $i=1,2,3$ at successive  non-dimensionalised time instants. The colour scale ranges from blue (corresponding to $0.5$) to red (corresponding to $1$). The black lines highlight the boundaries of the subdomains ${\cal D}_1$, ${\cal D}_2$ and ${\cal D}_3$, and the effective interfaces $\Sigma_{12}$ and $\Sigma_{23}$. The white curves are isolines that track the region of space occupied by the cancer multicellular mass. To construct numerical solutions, we impose zero Neumann boundary conditions on the outer boundaries of the subdomains for all dependent variables. We consider a biological scenario whereby cancer cells are initially confined to a circular region of the ovary centred at the point $\x_0$, while healthy cells occupy the rest of the spatial domain, and no MMPs are initially present. Hence, we assume $\rho_i(0,\x):=\rho_0 \; \text{ for all } \x \in {\cal D}_i$ and $i=\left\lbrace 1,2,3\right\rbrace$, 
$c(0,\x) \equiv 0$ and $\varphi(0, \x) := -1 + 2 \exp \left(|\x - \x_0|^2/b\right)$. Finally, we choose $\rho_0 = 0.5$, $\mu_1 = \mu_2 = \mu_3 \equiv\bar{\mu} = 0.5$, $\bar{\mu}_{23} = 1$, $\x_0=(-0.13,1.04)$, $b=0.01$, $K_c=0.2$, $D_c= 0.005$, $\gamma_c=0.5$ and ${\mu}_{12}(\x) := \bar{\mu}_{12} \, \mathbf{1}_{\Sigma_p}(\x)$, where $\bar{\mu}_{12}=0.1$ and $\mathbf{1}_{\Sigma_p}(\x)$ is a mollification of the indicator function of the set $\Sigma_p \subset \Sigma_{12}$.}\label{fig:ovary}
\end{figure}

The numerical results obtained are summarised by the plots in Fig.~\ref{fig:ovary}. As illustrated by these plots, which display the cell volume fraction in the different subdomains along with the boundaries of the cancer multicellular mass (white lines), the mathematical model defined by the effective interface problem~\eqref{P0} posed on the spatial domain of Fig.~\ref{fig:geometria_ovaio} and coupled with the transmission problem~\eqref{eq:MMP} can qualitatively reproduce the salient steps of the metastatic journey undertaken by an ovarian cancer multicellular mass. 
In summary, cancer cells are initially confined to the ovary region ${\cal D}_1$ [\emph{vid.} Fig.~\ref{fig:ovarya}], where they proliferate and grow into a multicellular mass. At later stages [\emph{vid.} Figs.~\ref{fig:ovaryb}-\ref{fig:ovaryd}], cancer cells break through the damaged part of the ovarian capsule $\Sigma_{p} \subset \Sigma_{12}$ and spread across the peritoneal region ${\cal D}_2$, until they reach the peritoneal lining $\Sigma_{23}$. From there, secreting MMPs, cancer cells create one focus of invasion [\emph{vid}. Fig.~\ref{fig:ovarye}], which enables the multicellular mass to squeeze through the peritoneal lining and form a secondary tumour in the healthy tissue adjacent to the peritoneum ${\cal D}_3$ -- \emph{vid}. Figs.~\ref{fig:ovaryf}-\ref{fig:ovaryh}. 

Note that the plots in Figs.~\ref{fig:ovarye}-\ref{fig:ovaryh} indicate that the size of the focus of invasion grows over time. This is due to the diffusion of MMPs secreted by cancer cells, which increase the local value of the effective mobility coefficient $\mu_{23}(t,\x)$ [\emph{cf}. the expression given by Eq.~\eqref{eq:mu23tilderis}]. Moreover, throughout the simulations one can verify that the cell volume fractions can become discontinuous not only in the portions of the ovarian capsule and of the peritoneal lining that are impermeable, but also in the permeable part of the ovarian capsule and at the focus of invasion in the peritoneal lining. 

\section{Conclusions and research perspectives}
We have developed a formal asymptotic method to mathematically address biological problems of cell invasion through thin membranes (\emph{i.e.} the basement membrane and other ECM barriers of small thickness). We have showed how, starting from an original transmission problem in which thin membranes are represented as finite regions of small thickness, one can obtain a limiting transmission problem where each membrane is replaced by an effective interface, and we derived a set of biophysically consistent interface conditions to close the limiting problem. 

\red{The approximation of a thin porous layer with an effective interface and a set of suitable transmission conditions is a simplifying approach that has attracted attention in a wide range of application fields -- \emph{e.g.} heat transfer problems~\cite{Ochoa}, flow simulations in porous media with immersed intersecting fractures~\cite{Scialo}, structural mechanical problems~\cite{bonnet2016effective, bellieud2016asymptotic}, flows through thin membranes for biological applications~\cite{Kleinhans} -- as it brings considerable modelling and computational benefits. From the modelling point of view, the main benefit lies in the fact that, by using this approximation, one does not need to develop a detailed model of the phenomena that occur inside the thin layer. From the computational point of view, such an approximation ensures a stark reduction of simulation time in the case of very thin layers, since it makes it possible to avoid the computational cost associated with the fine mesh required to produce accurate numerical results in the proximity of a thin layer, where sharp variations of the dependent variables can lead to the emergence of numerical instabilities. The price to pay for having a simpler and more computationally efficient model is the introduction of effective interface parameters, such as our ``effective mobility coefficient'', the estimation of which may require {\it ad hoc} experiments and extensive parameter fitting.}

The formal results obtained have been validated via numerical simulations showing that the relative error between the solutions to the original transmission problem and the solutions to the limiting problem vanishes when the thickness of the membranes tends to zero. Moreover, in order to show potential applications of our effective interface conditions, we have employed the limiting transmission problem to model cancer cell invasion through the basement membrane and the metastatic spread of ovarian carcinoma. 

Our work can be extended both from the analytical perspective and from the modelling point of view. From the analytical perspective, it would be interesting to provide a rigorous proof of the formal results established by Proposition~\ref{Th1}. From the modelling point of view, we would like to generalise the results presented in this paper to the case of multiple cell populations. Moreover, it would be interesting to understand how to develop further our formal method for deriving effective interface conditions to consider momentum-related equations different from Eq.~\eqref{def.v}\red{, in order to capture the visco-elasto-plastic behaviour of cellular aggregates, which is induced by the dynamic formation of bonds between cells and by the interaction between cells and the extracellular environment, and the active response of living aggregates.} 

Although the focus of this work has been on cancer invasion, cell penetration of thin membranes occurs also during development, immune surveillance and disease states other than cancer, such as fibrosis \cite{rowe2008breaching}. Hence, the effective interface conditions that we have derived can find fruitful application in a variety of research fields in the biological and medical sciences, including developmental biology and immunology.  

\appendix
\renewcommand{\thesection}{S.\arabic{section}}

\section*{Supplementary Material}

 \section{Numerical solutions to a one-dimensional problem illustrating the results of Proposition 3.1}
In order to  illustrate the formal results established by Proposition 3.1, we construct numerical solutions to a one-dimensional thin layer problem $\mathcal{P}_{\e}$ of mobility $\mu_{2\e} = \e \, \bar\mu_2$, where $\bar\mu_2 > 0$. We compare the solutions obtained for decreasing values of $\e$ with the numerical solutions of the corresponding effective interface problem ${\cal P}_0$ of effective mobility $\tilde\mu_{13} = \mu_{2\e} / \e \equiv \bar\mu_2$. Throughout this section we make use of the notation $\x = x$.

For the solution of the thin layer problem ${\cal P}_{\e}$ to converge to a stationary profile that is non-constant in $x$, we consider a somehow artificial scenario whereby the cells proliferate according to a logistic law with intrinsic growth rate $r >0$ in the subdomain ${\cal D}_{1 \e}$, whereas cell proliferation is balanced by natural death in the subdomains ${\cal D}_{2 \e}$ and ${\cal D}_{3  \e}$. Under these assumptions, letting $L>0$ be the thickness of the region represented as the subdomain ${\cal D}_{1 \e}$, we introduce the non-dimensionalised independent variables $\hat{t} = r \, t$ and $\hat{x} = x/L$ so that, dropping the carets from the non-dimensionalised quantities, we have
$$
{\cal D}_{1 \e} :=(-1,0), \quad {\cal D}_{2  \e} := (0,\e), \quad {\cal D}_{3  \e} := (\e,1)
$$
and
$$
\Gamma_{1 \e}(\rho_{1 \e}) := \left(1- \rho_{1 \e}\right) \rho_{1 \e}, \quad \Gamma_{2 \e}(\rho_{2 \e}) = \Gamma_{3 \e}(\rho_{3 \e}) \equiv 0.
$$
We assume the cell mobility coefficients in the subdomains ${\cal D}_{1 \e}$ and ${\cal D}_{3 \e}$ to have the same constant value, \emph{i.e.} 
$$
\mu_{1 \e} = \mu_{3 \e} \equiv \bar{\mu} \quad \text{with} \quad \bar{\mu}>0.
$$
Moreover, we use the following barotropic relation
\begin{equation}\label{1dd}
p \equiv f(\rho) \quad \text{with} \quad  f(\rho):= (\rho - \rho_0)_+ \quad  \text{and} \quad \quad 0 < \rho_0 < 1, 
\end{equation}
where $(\cdot)_+$ is the positive part of $(\cdot)$. We impose zero Neumann boundary condition on the left outer boundary, a Dirichlet boundary condition on the right outer boundary, and the following initial conditions
$$
\rho_{i \e}(0,x) := \rho_0 \; \text{ for all } x \in {\cal D}_{i \e}, \quad i=1,2,3.
$$
These initial conditions model a biological scenario where the cells are initially uniformly distributed in space at the stress-free state. 

Similarly, for the effective interface problem ${\cal P}_0$ we consider
$$
\tilde {\cal D}_{1} :=(-1,0), \quad \tilde {\cal D}_{3} := (0,1) 
$$
$$
\tilde \Gamma_1(\tilde \rho_{1}) := \left(1- \tilde \rho_{1}\right) \tilde \rho_{1}, \quad   \tilde \Gamma_3(\rho_{3}) \equiv 0, \quad \tilde \mu_{1} = \tilde \mu_{3} \equiv \bar{\mu}, \quad \tilde \mu_{13}(\x) \equiv \bar{\mu}_{2}.
$$
Furthermore, we use the barotropic relation~\eqref{1dd}. We impose zero Neumann boundary condition on the left outer boundary, a Dirichlet boundary condition on the right outer boundary, and the following initial conditions
$$
\tilde \rho_i(0,x) := \rho_0 \; \text{ for all } x \in \tilde {\cal D}_i, \quad i=1,3.
$$

We remark that we consider a scenario whereby the cell volume fraction at $t=0$ is equal to or greater than $\rho_0$ for all $\x$. Since $\rho_0<1$, under the above definitions of the growth rates $\Gamma_{i \e}$ and $\tilde \Gamma_i$ both the thin layer problem $\mathcal{P}_{\e}$ and the effective interface problem $\mathcal{P}_{0}$ are such that the cell volume fraction will be greater than or equal to $\rho_0$ for all $t \geq 0$. Under this scenario, the barotropic relation~\eqref{1dd} is such that Assumption~\ref{ass.p} is satisfied.

\begin{figure}[h!]
\centering
\subfigure[]{\includegraphics[width=0.46\textwidth]{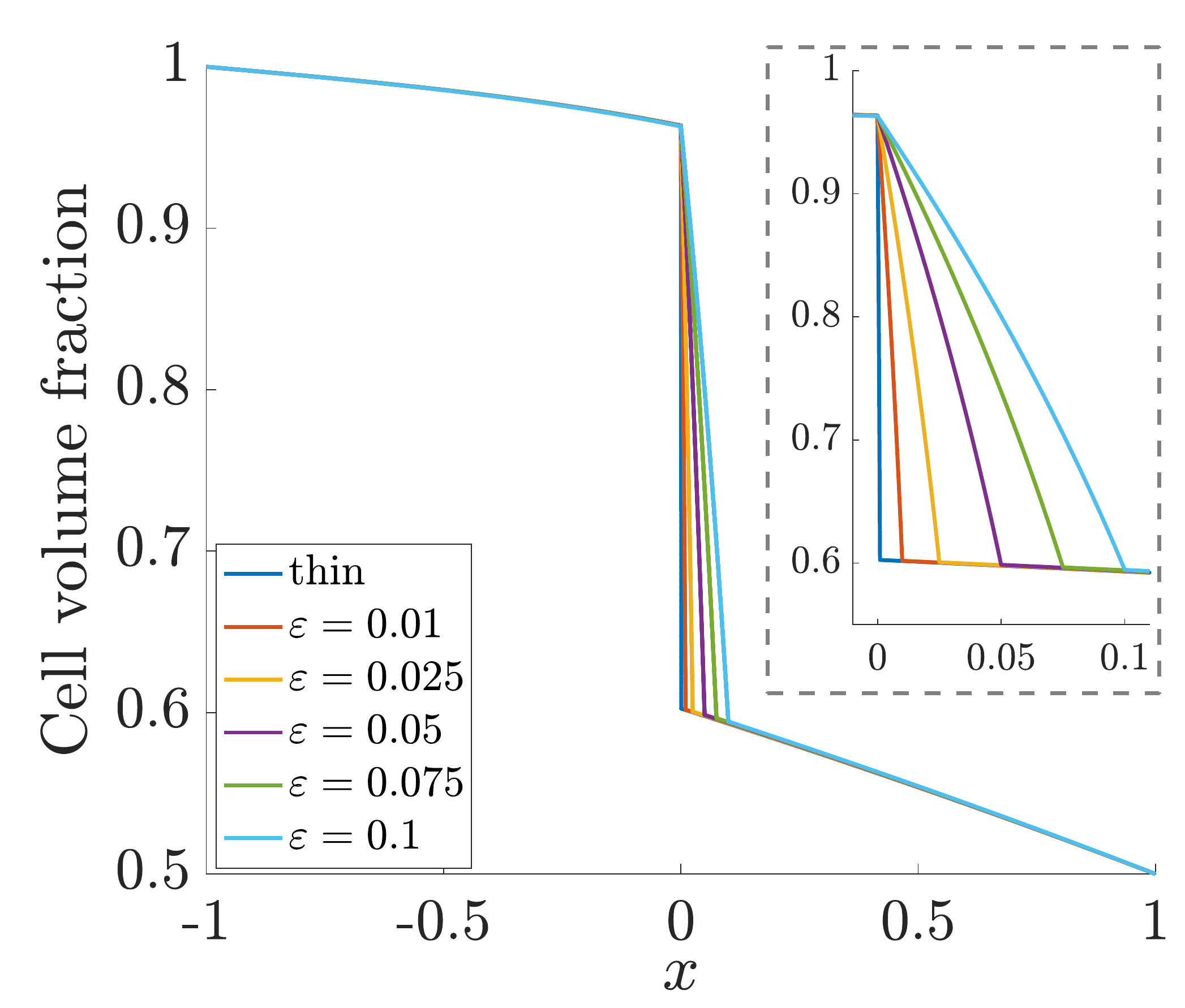}}
\subfigure[]{\includegraphics[width=0.46\textwidth]{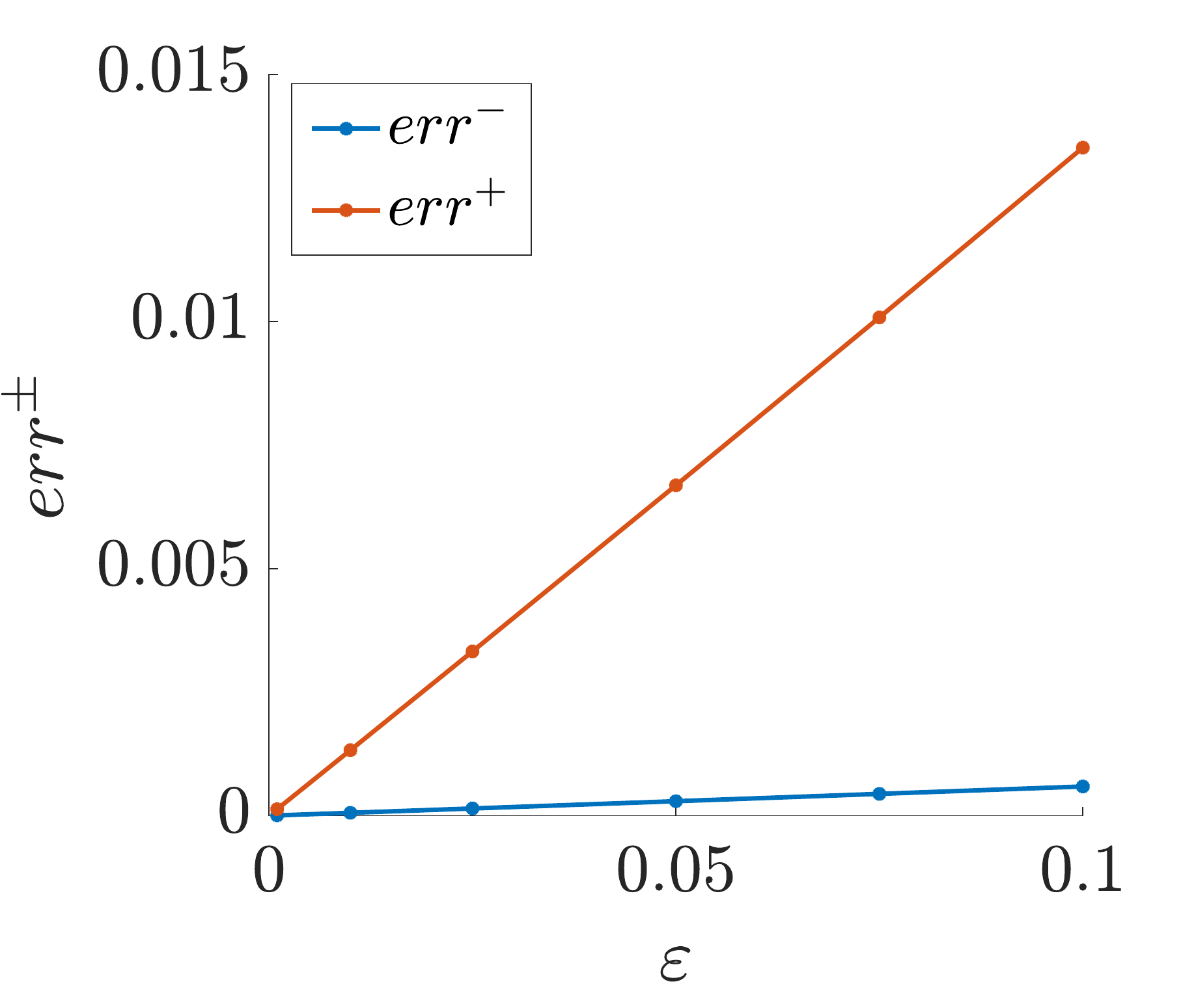}}
\caption{{\bf Numerical solutions to a one-dimensional problem illustrating the results of Proposition 3.1.} {\bf (a)} Comparison between the numerical solutions to the thin layer problem $\mathcal{P}_{\e}$ with thickness $\e~\in~ \{0.01, 0.025, 0.05, 0.75, 0.1\}$ and mobility $\mu_{2\e}=\e\bar\mu_2$ and the numerical solutions to the corresponding effective interface problem $\mathcal{P}_{0}$ with effective mobility $\tilde\mu_{13}~=~\bar\mu_2$ at time $t=20$, i.e. when the numerical solutions appear to be stationary. {\bf (b)} Relative error between the numerical solutions to the thin layer problem $\mathcal{P}_{\e}$ and the numerical solutions to the effective interface problem $\mathcal{P}_{0}$ as a function of $\e$ at the time instant $t=20$. The blue line displays the relative error $err^- = |\rho_{1 \e}(20,0) - \tilde \rho_1(20,0^-)| / \tilde \rho_1(20,0^-)$, while the red line displays the relative error $err^+ = | \rho_{3 \e}(20,\e) - \tilde \rho_3(20,0^+)| / \tilde \rho_3(20,0^+)$.}\label{fig:steady}
\end{figure}

To construct numerical solutions we choose 
\begin{equation}
\label{eqparval}
\rho_0 = 0.5, \quad \bar{\mu}=0.5 \quad \text{and} \quad \bar{\mu}_2 = 0.1
\end{equation}
and we carry out computational simulations for $t \in [0,20]$, since numerical solutions appear to be stationary at $t=20$. The results obtained are summarised by the plots in Fig.~\ref{fig:steady}, which display the numerical solutions to the effective interface problem $\mathcal{P}_{0}$ and the numerical solutions to the thin layer problem $\mathcal{P}_{\e}$ for decreasing values of $\e$ at $t=20$. 

The curves in Fig.~\ref{fig:steady}(a) indicate that the discrepancy between the numerical solutions to the thin layer problem and the numerical solutions to the effective interface problem decreases as $\e$ tends to zero. This is more precisely quantified by the curves in Fig.~\ref{fig:steady}(b), which display: the relative error between the numerical solutions to the thin layer problem at $x=0$ and the numerical solutions to the effective interface problem at $x=0^-$ for $t=20$ as a function of $\e$ (blue line); the relative error between the numerical solutions to the thin layer problem at $x=\e$ and the numerical solutions to the effective interface problem at $x=0^+$ for $t=20$ as a function of $\e$ (red line). In agreement with the formal results established by Proposition 3.1, both relative errors tend to zero (linearly) as $\e \to 0$. Taken together, the numerical results presented in this section illustrate that there is a good match between the numerical solutions to the original transmission problem with membrane thickness $\e$ and mobility $\mu_{2\e}$ and the numerical solutions to the limiting transmission problem with effective mobility $\tilde\mu_{13}$. Hence, when the thickness of the membrane represented as the subdomain ${\cal D}_{2 \e}$ is small, instead of solving the problem $\mathcal{P}_{\e}$ one can solve the approximate problem $\mathcal{P}_{0}$ with effective mobility coefficient $\tilde\mu_{13}$. The quality of the approximation is higher for membranes of smaller thickness.

\section{Formal derivation of the continuity condition (4.6) for $\tilde \varphi$}
Proceeding in the same way as at the beginning of the proof of Proposition 3.1, we rewrite Eq.~(4.3) for $\varphi_{\e}$ in ${\cal D}_{2 \e}$ as 
$$
\displaystyle{\frac{\partial \varphi_{\e}}{\partial t} - \mu_{2 \e} f'(\rho_{2\e}) \nabla_{\x_\Sigma} \rho_{2 \e} \cdot \nabla_{\x_\Sigma} \varphi_{\e} - \frac{1}{\e} \left(\frac{\mu_{2 \e}}{\e} f'(\rho_{2\e}) \frac{\partial  \rho_{2 \e}}{\partial \eta} \frac{\partial  \varphi_{\e}}{\partial \eta}\right)= 0.} 
$$
Multiplying both sides of the latter equation by $\rho_{2\e}$ and rearranging terms yields
$$
\displaystyle{\left(\frac{\mu_{2 \e}}{\e} \rho_{2\e} f'(\rho_{2\e}) \frac{\partial  \rho_{2 \e}}{\partial \eta}\right) \frac{\partial  \varphi_{\e}}{\partial \eta} = \e \rho_{2\e} \left(\frac{\partial \varphi_{\e}}{\partial t} - \mu_{2 \e} f'(\rho_{2\e}) \nabla_{\x_\Sigma} \rho_{2 \e} \cdot \nabla_{\x_\Sigma} \varphi_{\e} \right).} 
$$
Substituting the ansatz
$$
\varphi_{\e}\left(\dfrac{x_{\perp}}{\e}, \mathbf{x}_{\Sigma}\right) = \varphi^0(\eta, \mathbf{x}_{\Sigma}) + \varepsilon \varphi^1(\eta, \mathbf{x}_{\Sigma}) + \smallO(\varepsilon)
$$
and
$$
\rho_{2\varepsilon}\left(\dfrac{x_{\perp}}{\e}, \mathbf{x}_{\Sigma}\right) = \rho_2^0(\eta, \mathbf{x}_{\Sigma}) + \varepsilon \rho_2^1(\eta, \mathbf{x}_{\Sigma}) + \smallO(\varepsilon)
$$
into the above equation, letting $\e \to 0$  and using the fact that, as shown in the proof of Proposition~3.1, 
$$
\tilde \mu_{13} \, \rho^0_{2} \, f'(\rho^0_{2}) \, \frac{\partial  \rho^0_{2}}{\partial \eta} = \tilde \mu_1 \, \tilde \rho_1 \, f'(\tilde \rho_1) \, \nabla \tilde \rho_1 \cdot \tilde{\n}_{13} \big|_{\tilde \Sigma_{13}} = \tilde \mu_3 \, \tilde \rho_3 \, f'(\tilde \rho_3) \, \nabla \tilde \rho_3 \cdot \tilde{\n}_{13} \big|_{\tilde \Sigma_{13}} \quad \forall \eta \in (0,1)
$$
we formally obtain
$$
\frac{\partial  \varphi^0}{\partial \eta} = 0 \quad \Longrightarrow \quad \displaystyle{\varphi^0} = const. \;\; \forall \eta \in (0,1),
$$
from which we deduce the continuity condition (4.6) for $\tilde \varphi$. We remark that $\displaystyle{\tilde \mu_{13} \, \rho^0_{2} \, f'(\rho^0_{2}) \, \frac{\partial  \rho^0_{2}}{\partial \eta}}=0$ only if $\displaystyle{\tilde \mu_1 \, \tilde \rho_1 \, f'(\tilde \rho_1) \, \nabla \tilde \rho_1 \cdot \tilde{\n}_{13} \big|_{\tilde \Sigma_{13}}} = \tilde \mu_3 \, \tilde \rho_3 \, f'(\tilde \rho_3) \, \nabla \tilde \rho_3 \cdot \tilde{\n}_{13} \big|_{\tilde \Sigma_{13}}=0$. In this case, the cell flux across ${\tilde \Sigma_{13}}$ is identically zero and, therefore, the level set does not cross the effective interface.

\section{Derivation of the definition (4.7) of ${\mu}_{23}(t,\x)$}
Using a modelling strategy similar to that proposed by Gallinato \emph{et al.}~\cite{Gallinato} and Giverso \emph{et al.}~\cite{giverso2018nucleus}, we define the effective mobility coefficient ${\mu}_{23}$ as
\begin{equation} \label{eq:mu23tildeori}
{\mu}_{23}(t,\x) \equiv {\mu}_{23}(A(t,\x)) := \bar{\mu}_{23} \dfrac{ \left(A(t,\x) - A_0 \right)_+}{B + (A(t,\x)-A_0)}\, , \quad B>0,
\end{equation}
where $\bar{\mu}_{23}>0$ is the maximum mobility of ovarian cancer cells through the interface that models the peritoneal lining, the function $A(t,\x) > 0$ represents the average cross-section of the pores of the membrane at position $\x \in \Sigma_{23}$ and time $t \geq 0$, and the parameter $A_0 > 0$ is the critical value of the average pores' cross-section below which, according to ``the physical limit of cell migration''~\cite{wolf2013physical}, the membrane is completely impermeable to cancer cells. The evolution of the function $A(t,\x)$ is governed by the following differential equation
\begin{equation} \label{eq:pori}
\dfrac{\partial A}{\partial t} = \alpha \left(A_1- A\right)  + \beta \, c \qquad {\rm on}\ \Sigma_{23}.
\end{equation}
In Eq.~\eqref{eq:pori}, the parameter $A_1>0$, with $A_1<A_0$, models the non-dimensionalised average cross-section of the pores of the membrane in physiological conditions, $\alpha$ is the rate of remodelling of $A$ to the normal value $A_1$, and $\beta >0$ represents the rate at which MMPs increase the size of the pores of the membrane. Under the biologically realistic assumption that the process of pore cleavage and repairing is much faster than tumour expansion, we assume the average cross-section of the pores of the membrane to be in quasi-stationary equilibrium and rewrite Eq.~\eqref{eq:pori} as
$$
0 = \alpha \left(A_1- A(t,\x)\right)  + \beta \, c(t,\x) \quad \Longrightarrow \quad A(t,\x) = A_1 + \frac{\beta}{\alpha} \, c(t,\x).
$$
Substituting the above expression for $A(t,\x)$ into~\eqref{eq:mu23tildeori} and rearranging terms gives
\begin{equation} \label{eq:mu23tilderis}
{\mu}_{23}(t,\x) \equiv {\mu}_{23}(\hat{c}(t,\x)) = \bar{\mu}_{23} \dfrac{ \left(\hat{c}(t,\x) - 1 \right)_+}{K_c + (\hat{c}(t,\x)-1)}\, ,
\end{equation}
with
$$
\hat{c}(t,\x) = \frac{\beta}{\alpha (A_0-A_1)} \, c(t,\x) \quad \text{and} \quad K_c = \frac{B}{A_0-A_1} \, .
$$
In Section 4.2, we use~\eqref{eq:mu23tilderis} and, with a slight abuse of notation, we rename the rescaled concentration of MMPs $\hat{c}(t,\x)$ to $c(t,\x)$. 

\bibliographystyle{siam}
\bibliography{biblio}

\end{document}